\documentclass{jfm}

\usepackage{graphicx}
\usepackage{dcolumn}% Align table columns on decimal point
\usepackage{epstopdf, epsfig, amsmath, bm}
\usepackage{etoolbox}

\usepackage[utf8]{inputenc}
\usepackage[T1]{fontenc}

\usepackage{newtxtext}
\usepackage{newtxmath}
\usepackage{natbib}
\usepackage{hyperref}
\usepackage{comment}
\hypersetup{
    colorlinks = true,
    urlcolor   = blue,
    citecolor  = black,
}

\newcommand{\RomanNumeralCaps}[1]
\linenumbers

\newcommand{\be}{\begin{equation}}
\newcommand{\ee}{\end{equation}}
\newcommand{\bes}{\begin{equation*}}
\newcommand{\ees}{\end{equation*}}
\newcommand{\bea}{\begin{eqnarray}}
\newcommand{\eea}{\end{eqnarray}}
\newcommand{\bi}{\begin {itemize}}
\newcommand{\ei}{\end {itemize}}
\newcommand{\benm}{\begin{enumerate}}
\newcommand{\eenm}{\end{enumerate}}
\newcommand{\bmn}{\begin{minipage}}
\newcommand{\emn}{\end{minipage}}
\newcommand{\bfig}{\begin{figure}}
\newcommand{\efig}{\end{figure}}
\newcommand{\ig}{\includegraphics}
\newcommand{\lnw}{\linewidth}

\newcommand{\bcls}{\begin{columns}}
\newcommand{\ecls}{\end{columns}}
\newcommand{\bcl}{\begin{column}}
\newcommand{\ecl}{\end{column}}

\newcommand{\bmat}{\begin{matrix}}
\newcommand{\emat}{\end{matrix}}
\newcommand{\bpmat}{\begin{pmatrix}}
\newcommand{\epmat}{\end{pmatrix}}
\newcommand{\bvmat}{\begin{vmatrix}}
\newcommand{\evmat}{\end{vmatrix}}
\newcommand{\bbmat}{\begin{bmatrix}}
\newcommand{\ebmat}{\end{bmatrix}}

\newcommand{\ovl}{\overline}
\newcommand{\ol}{\overline}

\newcommand{\ptl}{\partial}
\newcommand{\td}{\widetilde}

\newcommand{\lal}{\langle}
\newcommand{\ral}{\rangle}

\newcommand{\ep}{\epsilon}

\renewcommand{\k}{{\bm k}}

\renewcommand{\u}{{\bm u}}

\newcommand{\f}{{\bm f}}

\newcommand{\x}{{\bm x}}

\newcommand{\obu}{{\ol{\bm u}}}

\newcommand{\obf}{{\ol{\bm f}}}

\newcommand{\olp}{{\ol{p}}}
\newcommand{\ols}{{\ol{s}}}
\newcommand{\olu}{{\ol{u}}}

\newcommand{\bu}{{\bm u}}
\newcommand{\bv}{{\bm v}}
\newcommand{\bx}{{\bm x}}

\newcommand{\EE}{{\mathcal E}}

\graphicspath{{./data/}}

\title{Intrinsic relationship between synchronisation thresholds and Lyapunov vectors: evidence from large eddy simulations and shell models}

\author{Jian Li\aff{1}, Wenwen Si\aff{1}, Yi
Li\aff{2}\corresp{\email{yili@sheffield.ac.uk}},
Peng Xu\aff{1}
} 
\affiliation{
  \aff{1}School of Naval Architecture and Maritime, Zhejiang Ocean
University, Zhoushan, 316022, China
  \aff{2}School of Mathematical and Physical Sciences, University of
  Sheffield, Sheffield, S3 7RH, UK 
}

\begin{document}
\maketitle

\begin{abstract}
  
An important parameter characterising the synchronisation of
turbulent flows is the threshold coupling wavenumber. 
This study investigates the relationship between the 
threshold coupling wavenumber and the leading Lyapunov vector
using
large eddy simulations and 
the SABRA model. Various subgrid-scale
stress models, Reynolds numbers, and different
coupling methods are examined. 
A new scaling relation is
identified for the leading Lyapunov exponents in large eddy
simulations, showing that they approximate those of filtered direct
numerical simulations. This interpretation provides a 
physical basis for results related to the Lyapunov exponents of large
eddy simulations, including those related to synchronisation. 
Synchronization experiments show 
that the peak wavenumber of the energy spectrum
of the leading Lyapunov vector coincides with the threshold
coupling wavenumber, in 
large eddy simulations of box turbulence with standard
Smagorinsky or dynamic mixed models as well as in the SABRA model,
replicating results from 
direct numerical simulations of box turbulence. 
Although the dynamic Smagorinsky model exhibit different behaviour,
the totality of the results 
suggests that the relationship is an intrinsic property of a
certain class
of chaotic systems. 
We also confirm that conditional Lyapunov exponents
characterize the synchronization process in indirectly coupled systems as they do in
directly coupled ones,
with their values insensitive to the nature of the master flow.
These findings advance the understanding of the role of 
the Lyapunov vector in the synchronization of turbulence.

\end{abstract}

\section{Introduction }\label{sect:intro}

Synchronisation of turbulent flows has been a topic of great interest in
recent years. In the simplest form, the technique seeks to reproduce the
evolution of one flow in another, by feeding incomplete data obtained from the
former into the latter through some coupling mechanism. The technique is at
the core of applications such as data assimilation
\citep{Kalnay03} where there is a need to enhance the control, construction 
or prediction of instantaneous turbulence
field. 
Though the interest in the topic sees resurgence recently, the earliest
research can be traced back to two decades ago. \citet{Henshawetal03} 
derives
theoretical estimate for the number of Fourier modes required to
synchronise two turbulent solutions of Burgers' equation and
those of unforced
Navier-Stokes equations (NSE). 
Numerical simulations are conducted which verifies the
theoretical estimate but also show the latter tends to
over estimate the bound. Soon after, \citet{Yoshidaetal05}
investigate the complete synchronisation of two isotropic turbulent flows in
three-dimensional (3D) periodic boxes and find that it is achieved if all
Fourier modes with wavenumber less than at least $k_c$ are
coupled, where $k_c \eta\approx 0.2$ and $\eta$ is the Kolmogorov length scale
(see also \citet{Lalescuetal13}). More recent work on the synchronisation
of two direct numerical simulations (DNS) of isotropic turbulence
includes those using the
nudging coupling \citep{Leonietal18, Leonietal20} and the investigation on 
partial synchronisation of intense vortices when the number of
coupled modes is smaller than the threshold value
\citep{VelaMartin21}. 
\citet{NikolaidisIoannou22} investigates the synchronisation of Couette flows
by coupling selected streamwise Fourier modes. They compute the conditional leading
Lyapunov exponents (LLEs) \citep{Boccalettietal02} and show that
synchronisation happens when the conditional LLE is negative. Channel flows are
investigated by \citet{WangZaki22} where the synchronisability of the flows
by coupling velocity from different spatial domains is
investigated. This work appears to be the first to investigate 
coupling in physical space. The new scaling relation of the coupling domain is
obtained. The behaviour of synchronisation is also investigated when the coupling is applied at the
interface between two domains. \citet{Inubushietal2023} revisit the
synchronisation of turbulence in a periodic box at Reynolds
numbers significantly higher
than attempted before. Though they also find $k_c \eta$ to be approximately
$0.2$ based on the results for the conditional LLEs (the transverse Lyapunov exponents in
their terminology), it is noted that the conditional LLEs, hence the
exact threshold value for $k_c \eta$, depend on the Reynolds number. 
\citet{Wangetal23} investigate the cases when the
large-scale data is not sufficient to dominate the small scales. Several 
techniques are employed to 
improve the correlation between the small scales when only partial
synchronisation is achieved. Using nudging coupling, 
\citet{BuzzicottiLeoni20}  
identify the optimal parameters in 
subgrid-scale (SGS) stress models that minimise the
synchronisation error between synchronised LES and DNS.   
\citet{Lietal22} study the synchronisation between LES and DNS using the
master-slave coupling, focusing on the threshold wavenumber and the behaviours
of different SGS stress models.  
Most recently, \citet{Lietal24} reports an investigation into
the synchronisation of rotating turbulence in a 3D periodic box. They show
that the forcing scheme has significant impacts on
the threshold wavenumber $k_c$. 
More interestingly, they find that the energy spectra of
the leading Lyapunov vectors (LLVs) have peaks which always
locate approximately at the
threshold wavenumber $k_c$, 
regardless the rotation rates and the forcing mechanisms.

The present research is mainly concerned with the last observation
mentioned above, namely there appears to be an intrinsic relationship
between the peak in the energy spectrum of the LLV and the
threshold wavenumber for synchronisation. 
Though \citet{Lietal24} are the first to highlight the link 
between the two wavenumbers, they are not the first to
investigate the energy spectrum of the LLV. In
\citet{BudanurKantz22}, the energy spectrum of the LLV for
turbulence in a periodic box
is calculated for simulations with up to $432^3$ grid points. The
spectrum is found to peak at $k \eta \approx 0.2$ at all Reynolds
numbers.
The energy spectrum of small velocity perturbation in box
turbulence
is investigated by \citet{Geetal23} with up to $512^3$ grid
points. When 
the perturbation experiences self-similar growth (at which stage it is 
analogous to 
the LLV), the peak of its energy spectrum is found to locate at $kL_\Delta
\approx 2$ where $L_\Delta$ is the integral length scale of the
perturbation. It can be deduced from the data in the paper 
that this also corresponds to $k\eta \approx
0.2$. In other words, the peak wavenumbers found in these two
works are both the same as the
threshold wavenumber for synchronisation for box turbulence.
In rotating turbulence, \citet{Lietal24} show that the two wavenumbers display similar dependence on the
rotation rate. In particular, for flows with constant power
forcing, both decrease as rotation rate is increased. 
Furthermore, there also seems to be a relationship between the
synchronisation threshold and the peaks of the LLV in Couette
flow. \citet{NikolaidisIoannou22} 
note in passing that the threshold wavenumber they find coincides with the peak of 
the spectra of both the vertical and span-wise velocity components
at a distance of $14$ wall units from the wall
for the LLV reported in
\citet{Nikitin2018}. The authors stop short of suggesting that
there is an intrinsic link between the two parameters.

The evidence summarised above is highly suggestive. 
However, an analytical theory explaining this relationship is yet to be developed. 
Therefore, one may justifiably question whether it 
is only coincidental, and further numerical evidence is needed to 
substantiate the relationship. 
One obvious approach is to demonstrate this
relationship in DNS with much higher Reynolds numbers. 
However, arguably the evidence is more convincing 
if it shows that this
relationship also exists in other chaotic systems, thus
indicating
the observation in DNS turbulence is an example of a more
general principle. 
Guided by this view, we address in this paper the question of whether the
relationship is genuine 
with numerical
experiments for the synchronisation between LES (rather than DNS), and supplement
these experiments with synchronisation experiments based on 
shell models. Specifically, we seek to 
show that the peak wavenumber of the energy spectrum
of the LLV and the threshold wavenumber for synchronisation are
the same for both LES turbulence and shell models. We contend
that the positive
outcome from these experiments will serve to substantiate 
the authenticity of the relationship in DNS and real turbulence. 
It is worth emphasising that we do not expect, and do not intend
to show, that the threshold or the peak
wavenumbers in LES or shell models are the same as
their counterparts in the DNS of box turbulence. 
In essence, LES is not chosen as a model for DNS, but rather a
nonlinear dynamical system which is different from DNS. 
Nevertheless, given the widespread use of LES in turbulent
modelling, the findings of this investigation may have yet to be seen implications for
turbulent simulations too.  

Shell models are a class of 
reduced order models which capture important features of turbulence.
A detailed discussion about shell models can be found in the monograph
by \citet{Ditlevsen11} (see also \citet{Biferale03} and
\cite{Bohretal98}). 
The shell model we use in present research is the SABRA model.
The state of the model is described by a vector of complex
numbers, which
mimics the Fourier modes of turbulent velocity. The system evolves 
according to a set of coupled nonlinear ordinary differential equations. 
Two copies of the system can be
synchronised in a way similar to the synchronisation of two DNS
or two LES, by coupling some components of the state vectors
together during the evolution. Therefore we may investigate and
characterise the
synchronisation of the SABRA model in a similar way. On the other hand, the simplicity of the
SABRA model allows one to simulate 
a much wider range of length scales that are not reachable even
with LES, and that is not interfered by the SGS model. These
benefits prompt us to supplement LES with
simulations of the SABRA model. As we will
show later, the SABRA model does display
features that are not present in LES. 

To examine the relationship between the LLV and the threshold wavenumber more fully,
the synchronisation experiments 
for LES are conducted for three different SGS models: the standard
Smagorinsky model, the dynamic Smagorinsky model, and the dynamic
mixed model. Our hope is to show that
the relationship is universally observed for different SGS
stress models. However, as will be explained below, the
picture is more complicated than this, with the DSM exhibiting
somewhat different behaviours. 

In the course of addressing the main objective laid out above, it becomes
clear that two auxiliary questions need to be answered so that
we can interpret the main results of the investigation in a meaningful way. The first
one is on the interpretation of the LLE of an LES. 
The LLE of an LES has been calculated previously \citep{Nastacetal17, BudanurKantz22}. The general
observation is that the LLE of an LES is lower than that of the DNS of the
same flow, and that it decreases as the filter length scale increases. 
However, the physics behind
these results is unclear. We will address this question as
the second objective of current research. We report a scaling
law for the LLE
of an LES that has not been discussed before, and show that the
LLE of an LES should be understood as an approximation to the LLE of
a filtered DNS. Though this result is auxiliary to our
main objective, we consider 
it a main result of this paper given its foundational nature. 

The second auxiliary question is the relationship between the synchronisation
threshold and the master flow. This question arises because
in practice the master system one uses to drive 
synchronisation is usually different from the slave
system. For example, in both \citet{BuzzicottiLeoni20} and \citet{Lietal22}
the slave is an LES whereas the master is a DNS. How
the threshold and the conditional LLEs depend on the
master system is a question one needs to answer in order to
gauge if the results we obtain will be relevant to different practices. We
address this question as the third objective of this paper. We do
so by comparing synchronisation experiments with
different master--slave coupling, which will be described below.
We present the results related to this objective in the appendices.

This paper documents our attempt at answering the above
questions. We start with a summary of the governing equations in
Section \ref{sect:eq}, where we also explain the various coupling
used in the synchronisation experiments and the definitions of
the conditional LLEs in the context of LES and the SABRA model.  
The details of the numerical experiments are explained in
Section \ref{sect:sim}. The results are presented and discussed
in Section \ref{sect:rd}. 
The results related to the first auxiliary question above are
considered fundamental for the interpretation of results related to the LLEs of LES,
therefore are presented first in this section. We then focus
on the relationship between the peak wavenumbers for the LLVs and
the threshold wavenumber in the synchronisation for LES. The
results for the SABRA model are discussed next. 
The conclusions are summarised in Section \ref{sect:con}. 
The results related to the second auxiliary question are
summarised in the appendices, as a complement to the main findings of the
investigation.

%TODO: some ideas - seems 2D turbulence has not been done too much. Boffeta
%did something about 2D finite size Lyapunov exponents. Daniel Clark did some
%on 2D turbulent in general including Kolmogorov - Sinai entropy etc. But
%still synchronisation is different? 
\section{Governing equations \label{sect:eq}}

\subsection{Flow field simulations}
We consider incompressible turbulent flows in a 3D periodic box $[0,
2\pi]^3$. DNS is conducted for flows with relatively low Reynolds numbers, where 
we integrate the Navier-Stokes equations (NSE)
\be \label{eq:nse}
\ptl_t \bu + (\bu \cdot \nabla) \bu = - \nabla p + \nu \nabla^2 \bu + \f,
\ee
in which $\bu \equiv (u_1, u_2, u_3)$ is the velocity field, $p$ is the 
pressure divided by the constant density, $\nu$ is the viscosity, and $\f$
is the forcing term. The velocity is assumed to be incompressible so that
\be \nabla \cdot \bu = 0. 
\ee
LES is conducted for flows with high Reynolds numbers, which is
based on the filtered NSE (fNSE)
\be \label{eq:fnse}
\ptl_t \obu + (\obu \cdot \nabla) \obu = - \nabla \olp+ 
\nabla \cdot (-\bm\tau) + \nu \nabla^2 \obu + \obf + \f_s,
\ee
and the filtered continuity equation $\nabla \cdot \obu =
0$. In these equations, the line $\ovl{\phantom{a}}$ represents filtering with filter
length $\Delta$. Therefore, $\obu$ and $\olp$ are the filtered velocity and
pressure, respectively. The SGS stress tensor $\bm\tau$ in Eq. (\ref{eq:fnse}) is defined as $\tau_{ij} = \ovl{u_iu}_j - \ovl{u}_i \ovl{u}_j,$
whereas $\obf$ is the filtered forcing term and $\f_s$ symbolically
represents the effects of coupling in synchronisation experiments, which will
be further explained below.  

We let $\f = (a_f\cos k_f x_2,0,0)$
with  $a_f = 0.15$ and $k_f = 1$ in all simulations. 
The flow resulting from this type of deterministic sinusoidal forcing is often called the Kolmogorov flow
\citep{BorueOrszag96}. The flow is inhomogeneous with a sinusoidal mean velocity
profile. Previous research
\citep{Yoshidaetal05, Inubushietal2023, Lietal24} shows that
the forcing only has at most secondary effects on the synchronisability of
flows in periodic boxes for non-rotating turbulence. Therefore, 
the conclusions we draw can be compared with previous research
where different forcing is used.

Simulations with different SGS stress models are conducted to 
assess the reproducibility of key observations.
Three canonical models are considered, namely the standard
Smagorinsky model (SSM), the dynamic Smagorinsky model (DSM),
and the dynamic mixed model (DMM). Details of these models can be found
in, 
e.g., \citet{MeneveauKatz00}. We summarise the formulae here for references.
The SSM is defined with 
\be \label{eq:ssm}
\tau_{ij} = - 2(c_s \Delta)^2 \vert\ovl{\bm s}\vert \ols_{ij}, 
\ee
where $c_s = 0.16$ is the Smagorinsky coefficient, $\ols_{ij}$ is the filtered strain rate 
tensor given by
\be
\ols_{ij} = \frac{1}{2}(\ptl_j \olu_i + \ptl_i \olu_j),
\ee
and $\vert \ovl{\bm s}\vert\equiv (2 \ols_{ij}\ols_{ij})^{1/2}$ is the
magnitude of the filtered strain rate. The DSM is also defined by Eq.
(\ref{eq:ssm}), but the coefficient $c_s^2$ is calculated dynamically using
the Germano identity \citep{Germano92}. 
Following the procedure as described in, e.g., 
\citet{Germano92, MeneveauKatz00}, one can derive  
\be \label{eq:cs2dyn}
c_s^2 = \frac{\lal L_{ij} M_{ij}\ral_v}{\lal M_{ij} M_{ij}\ral_v}, 
\ee
with 
\be
M_{ij} = -2 \Delta^2 \left[4 \vert \td{\ovl{\bm s}} 
\vert \td{\ols}_{ij} - \td{\vert \ovl{\bm s}\vert \ols_{ij}}\right], \quad
L_{ij} = \td{\olu_i \olu_j} - \td{\olu}_i \td{\olu}_j,
\ee
where $\lal \phantom{~}\ral_v$ represents volume average, and
$\td{\phantom{u}}$ denotes test-filtering with filter 
scale $2\Delta$ using the cut-off filter \citep{Pope00}. 
The model expression for DMM is given by 
\be
\tau_{ij} =  - 2(c_s \Delta)^2 \vert\ovl{\bm s}\vert \ols_{ij} 
+ c_{nl}\Delta^2 \ovl{A}_{ik} \ovl{A}_{jk} ~,
\ee
where $\ovl{A}_{ij} \equiv \ptl_j \ovl{u}_i $ is the filtered velocity
gradient and $c_{nl}$ is the nonlinear coefficient which is to be determined
dynamically.
Following the dynamic procedure, 
the expressions for $c_s^2$ and $c_{nl}$
are found to be
\begin{align}
& c_s^2 = \frac{\lal L_{ij} M_{ij} \ral_v \lal N_{ij} N_{ij} \ral_v - \lal L_{ij} N_{ij} \ral_v \lal
M_{ij} N_{ij}\ral_v}{\lal M_{ij} M_{ij} \ral_v \lal N_{ij} N_{ij} \ral_v - 
\lal M_{ij} N_{ij} \ral_v^2 },
\\
& c_{nl} = \frac{\lal L_{ij} N_{ij} \ral_v \lal M_{ij} M_{ij} \ral_v - \lal L_{ij} M_{ij} \ral_v 
\lal
M_{ij} N_{ij}\ral_v}{\lal M_{ij} M_{ij} \ral_v \lal N_{ij} N_{ij} \ral_v - 
\lal M_{ij} N_{ij} \ral_v^2 },
\end{align}
where 
\be
N_{ij} = \Delta^2 \left[4 \td{\ovl{A}}_{ik}\td{\ovl{A}}_{jk} - \td{\ovl{A}_{ik}
\ovl{A}}_{jk}\right].
\ee	
The three models are compared in depth in \citet{Kangetal03} with experimental
data. DMM produces better agreement in most
statistics and correlates better with the real SGS stress. SSM is known to be
too dissipative. DSM has the same expression as SSM, but $c_s^2$, obtained dynamically, 
fluctuates slightly over time and in general is somewhat smaller. As a
consequence, DSM is
less dissipative compared with SSM, and in general, agrees with experimental
data better than SSM. 

\subsection{The coupling between the masters and slaves \label{sect:coupling}}
\bfig
\centering
\ig[width = 0.6\lnw]{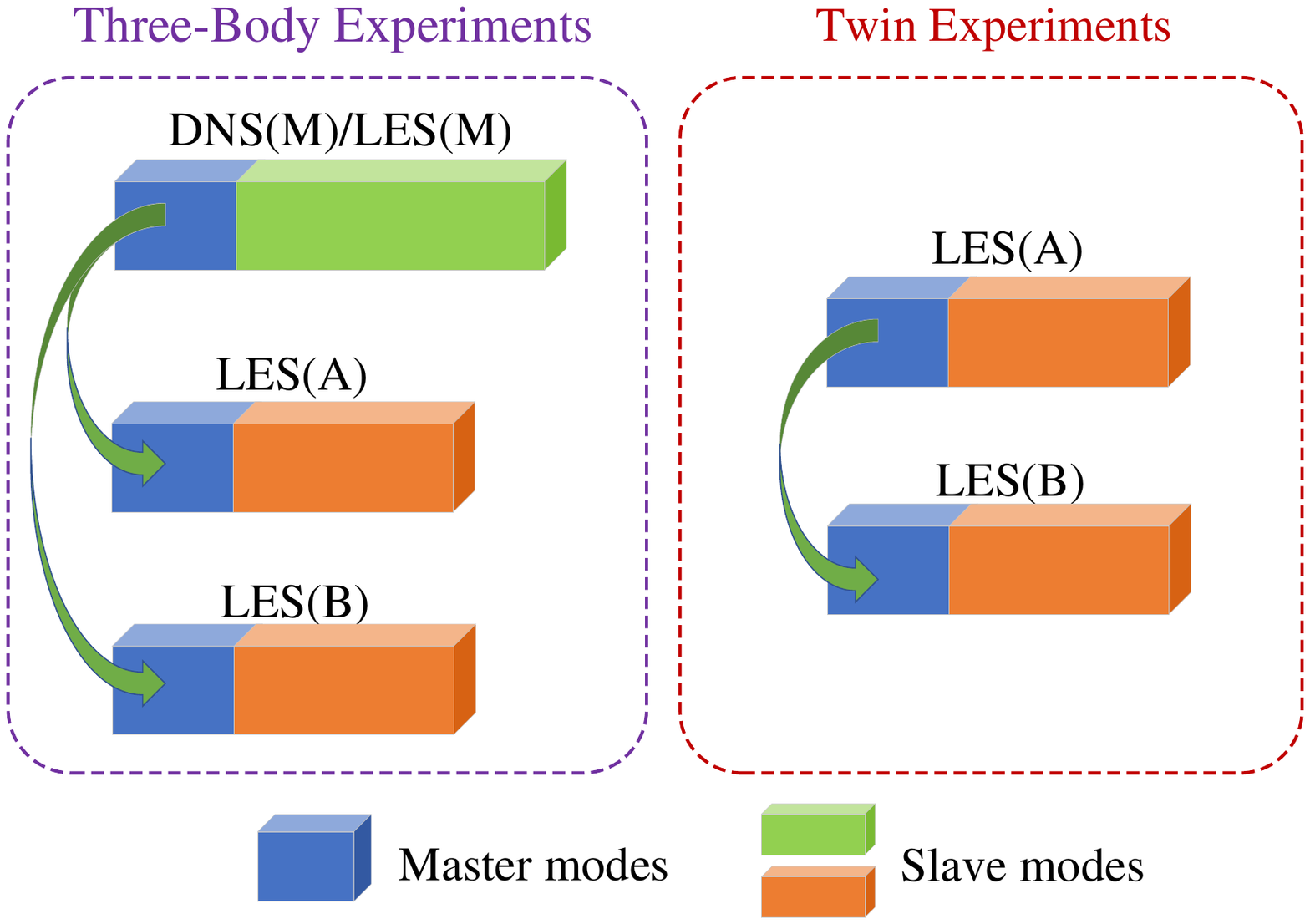}
\caption{\label{fig:cases} The setup of the
three-body experiments and the twin experiments showing differing coupling
mechanisms. The master in the former is a DNS in
Group I and an LES in
group II. It may have a different number of 
slave modes from the slave
systems, and its slave modes do not converge to those
in the slave systems $A$ and $B$.}
\efig

We run two groups of synchronisation experiments, and the
experiments in each
group are categorised into two sets with differing coupling methods.
The setup of the experiments is illustrated in Fig.
\ref{fig:cases}. 
Group I consists of numerical
experiments with lower Reynolds numbers so that the results can
be compared with DNS. Group II involves simulations with higher Reynolds
numbers that are not reachable by DNS. Therefore, DNS is
replaced by LES in Group II. 
In the first set of experiments,
a master system $M$ and two slave systems $A$ and $B$ are simulated
concurrently. We thus call them three-body experiments.
In three-body experiments, systems $A$ and $B$ are identical LES with the same SGS stress
model, but they have different initial conditions. The two systems are synchronised indirectly through their coupling
with the same master $M$. System 
$M$ is a DNS in Group I, but it is an LES in Group II 
that has a SGS stress model different from the one in systems $A$ and $B$. 
In the second set of experiments, systems $A$ and $B$ are coupled
directly together, with $A$ now serving as the master. These
experiments are called twin experiments.  

The three-body experiments are similar to those 
reported by \citet{Lietal22}. However, for such three-body
experiments, the properties of the conditional LLEs and the
LLVs, and their relationships with the synchronisation threshold
have not been investigated, which are 
address here.  
The twin experiments are similar to those reported in, e.g.,
\citet{Yoshidaetal05, Inubushietal2023, Lietal24}, in the sense that they all deal
with synchronisation between identical systems. However, here the
systems are LES as opposed to DNS. How the conditional LLEs and the
LLVs behave differently is the question that motivates this set of
experiments. 
Further more, the contrast between the two sets of experiments 
allows us to assess the impacts of the coupling mechanism: indirect
coupling via a third (different) master system versus
direct coupling. Note that the results addressing the questions outlined
above are mostly summarised in the appendices. In the
main text the discussion centres around the relationship between
the threshold wavenumber and the LLV, using the data gathered
from both the three-body and the twin experiments.

We now provide the mathematical representation of the coupling
and introduce several definitions. 
We use $\bu_{M}(\bx,t)$ to denote the velocity of system $M$, and $\hat{\bu}_{M}(\k,t)$
its Fourier mode, where $\k$ is the wavenumber. Similarly, the
duos $\obu_{A}$ and
$\hat{\obu}_{A}$, and  $\obu_{B}$ and $\hat{\obu}_{B}$, are used
to denote the velocity and its Fourier modes for 
systems $A$ and $B$, respectively. 
We use $\vert \k\vert$ to
denote the Euclidean norm of $\k$, i.e., $\vert \k \vert \equiv (k_1^2 + k_2^2 +
k_3^2)^{1/2}$.  

Let $k_m$ be a constant, and suppose we intend to couple the
Fourier modes with wavenumbers $\vert \k \vert \le k_m$. 
In the three-body experiments, the coupling is implemented by
letting
\be \label{eq:abm}
\hat{\obu}_{A}(\k,t) =  \hat{\bu}_{M}(\k,t), \qquad\hat{\obu}_{B}(\k,t) = \hat{\bu}_{M}(\k,t), 
\ee 
for $\vert \k \vert \le k_m$ at each time step. 
The effects of Eq. (\ref{eq:abm}) are represented by $\f_s$ in
Eq. (\ref{eq:fnse}) for both systems $A$ and $B$.
In the twin experiments, the coupling is implemented by letting  
\be \label{eq:ab}
    \hat{\obu}_{B}(\k,t) = \hat{\obu}_{A}(\k,t) 
\ee 
for $\vert \k \vert \le k_m$. This coupling 
introduces the forcing $\f_s$ into Eq. (\ref{eq:fnse}) for system
$B$, but does not impact system $A$. 

The constant $k_m$ is called the coupling wavenumber.
The Fourier modes  with $\vert \k
\vert \le k_m$ are called the master modes whereas those with $\vert \k \vert
> k_m$ are called the slave modes.  
Systems $A$ and $B$ synchronise asymptotically when $k_m$ is
sufficiently large, i.e.,  the difference between $\bu^A$ and
$\bu^B$ decays towards zero over time. The smallest $k_m$ value for which such synchronisation
happens is the threshold wavenumber, denoted as $k_c$. 
The wavenumber $k_c$ might be different for different
experiments. The dependence of $k_c$ on the coupling, flow
parameters, and its relationship with the LLV is the main focus
of this paper.   

\subsection{Lyapunov exponents and Lyapunov vectors}

The synchronisation experiments are analysed via the LLEs (i.e., the leading Lyapunov
exponents), the conditional LLEs, and
the LLVs (i.e., the leading Lyapunov vectors) of the LES of the
flows. We introduce the relevant definitions in this subsection. 

As we have mentioned in Introduction, a physical interpretation
of the LLE of an LES is lacking, and we will show
that it should be interpreted as an approximation to the LLE of filtered DNS of
the same flow. Therefore we start with the definition of the
LLE of filtered DNS. 
Let $\bu$ be a DNS velocity, and $\bu_\Delta^\delta$ be the
filtered infinitesimal perturbation, which is zero for wavenumber
larger than the cut-off wavenumber $k_\Delta \equiv \pi/\Delta$ where $\Delta$ is 
the filter length. 
The Fourier mode of $\bu_\Delta^\delta$ is denoted by
$\hat{\bu}_\Delta^\delta$. By definition
\be \label{eq:filpert}
\hat{\bu}_\Delta^\delta (\k, t) = 0
\ee
for $\vert \k \vert > k_\Delta$. 
The perturbation $\bu_\Delta^\delta$ is divergence free and obeys the linearised NSE: 
\be \label{eq:uf}
\ptl_t \bu_\Delta^\delta + (\bu \cdot \nabla) \bu_\Delta^\delta + (\bu_\Delta^\delta \cdot
\nabla) \bu = - \nabla 
p^\delta + \nu \nabla^2 \bu_\Delta^\delta + \f^\delta,
\ee
where $p^\delta$ is the pressure perturbation and $\f^\delta$ is the
perturbation to the forcing term.
The LLE for the filtered DNS, denoted by $\lambda_\Delta$, is
defined as 
\be \label{eq:lmdd}
\lambda_\Delta = \lim_{t\to \infty} 
\frac{1}{t}\log \frac{\Vert \bu_\Delta^\delta(\x, t+t_0)\Vert}{\Vert
\bu_\Delta^\delta (\x, t_0)\Vert},
\ee
where $t_0$ is an arbitrary initial time 
and $\Vert \cdot \Vert$ represents
the 2-norm defined by (for a generic vector field $\bm w$) 
\be
\Vert \bm w \Vert \equiv  \lal \bm w \cdot \bm w \ral_v^{1/2}.
\ee
Obviously $\lambda_\Delta$ depends on the filter scale $\Delta$. It measures
the average growth rate of the perturbations only applied to the Fourier modes of
$\bu$ with length scales larger than $\Delta$. 

The conditional LLE of a slave 
system in a synchronisation experiment is defined in a similar
way. The conditional LLE measures the averaged growth rate of an infinitesimal 
perturbation to the slave modes of a slave system. Let $\obu$ be
the velocity of a slave, i.e., systems $A$ or
$B$ in a three-body experiment or system $B$ in a twin
experiment. 
We let $\obu^\delta$ be an
infinitesimal perturbation to \emph{the slave modes} of $\obu$. As the master
modes are not perturbed, we have 
\be \label{eq:cle0}
\hat{\obu}^\delta(\k,t) = 0 \quad \text{for}\quad \vert \k \vert \le k_m. 
\ee
By definition, the perturbation $\obu^\delta$ is solenoidal and obeys the linearised fNSE 
\be \label{eq:udel}
\ptl_t \obu^\delta + (\obu \cdot \nabla) \obu^\delta + (\obu^\delta \cdot \nabla) \obu = - \nabla 
\olp^\delta + \nabla \cdot (-\bm\tau^\delta) + \nu \nabla^2 \obu^\delta + \obf^\delta,
\ee
where $\olp^\delta$, $\obf^\delta$ and $\bm\tau^\delta$ are 
the pressure perturbation, the perturbation to the filtered forcing, and the
perturbation to the SGS stress $\bm \tau$, respectively. $\obf^\delta$ is
included for completeness, but it is actually always zero for the
forcing we are using. 	

The conditional LLE for the slave is defined in terms of $\obu^\delta$ in a way
similar to Eq. (\ref{eq:lmdd}). Let $\lambda(k_m)$ denote the
conditional LLE with coupling wavenumber $k_m$. 
Then $\lambda(k_m)$ is defined by \citep{Boccalettietal02,
NikolaidisIoannou22, Inubushietal2023} 
\be \label{eq:lyadef}
\lambda(k_m) = \lim_{t\to \infty} 
\frac{1}{t}\log \frac{\Vert \obu^\delta(\x, t+t_0)\Vert}{\Vert \obu^\delta
(\x, t_0)\Vert}. 
\ee
The conditional LLE $\lambda(k_m)$ is a function of $k_m$. It
becomes 
the usual (unconditional) LLE when $k_m=0$. 
Synchronisation is successful only when the conditional LLE is
negative, as is shown previously in twin experiments for
turbulent channel flows
\citep{NikolaidisIoannou22}, turbulence in periodic boxes
\citep{Inubushietal2023}, and rotating turbulence \citep{Lietal24}. 
As we show in the appendices, the observation also holds for the 
three-body experiments. 
Therefore, the threshold wavenumber $k_c$ for $k_m$ 
can be found from $\lambda(k_m)$ as the root of the
equation $\lambda(k_m) = 0$. 

For sufficiently large $t$ 
the perturbation velocity $\obu^\delta$ approaches the
conditional LLV or the (unconditional) LLV if $k_m=0$
\citep{Bohretal98,
KuptsovParlitz12, Nikitin2018}.
The conditional LLV represents the most unstable perturbation
to the slave modes on long-time average. 
The properties of the (unconditional) LLV for a turbulent channel
flow are analysed in \citet{Nikitin2018}. In our previous work
\citep{Lietal24}, we find that the threshold coupling wavenumber $k_c$
correlates closely with the peak of the mean energy spectrum of
the LLVs. One of the objectives of this work is to look into the
properties of the LLVs in relation to the synchronisability of LES. 

\subsection{The SABRA model}

The investigation based on LES is supplemented with experiments
conducted with the SABRA model. 
The state of the SABRA model is a vector ${\bm z}\equiv (u_1,
u_2, ..., u_{N_\textrm{SB}})$, where $u_n$ is a complex number
referred to as the $n$th shell velocity and $N_\textrm{SB}$ is
the number of shells in the model. 
The equation for $u_n$ ($n=1,2,..., N_\textrm{SB}$) reads 
\be \label{eq:sb}
\frac{du_n}{dt} + \nu k_n^2 u_n = i(k_n u_{n+2} u^*_{n+1} - b k_{n-1}
u_{n+1} u^*_{n-1} - c k_{n-2} u_{n-1} u_{n-2}) + f \delta_{n1}, 
\ee
where $\nu$ is the viscosity, $i$ is the imaginary unit, 
$k_n$ is referred to as the wavenumber for the $n$th shell, $^*$ denotes complex conjugate, 
$b$
and $c\equiv b-1$ are constant parameters, $f$ is the forcing
amplitude, and $\delta_{n1}$ is the Kronecker delta $\delta_{ij}$
with $(i,j)=(n, 1)$ which
indicates that the forcing is applied only to the first shell. 

We use the following standard values for the parameters. The wavenumber $k_n$
is given by $ k_n= k_0 q^n$ with $k_0 = 0.05$ and $q = 2$. The forcing
amplitude $f = (0.005, 0.005)$.  
We let $b = 1/2$ so that the system has a helicity-like inviscid invariant. 
This leads to $c = -1/2$ due to the relation $c = b-1$, which guarantees 
that the total energy of the system is also an inviscid invariant of the model. The
viscosity $\nu$ determines the Reynolds number, which will be
adjusted in the synchronisation experiments. 

For suitable parameters, the evolution of SABRA model
becomes
chaotic. In this regime, the energy spectrum of the
system displays an inertial
range where a $k_n^{-5/3}$ distribution is observed. That is, the slope of
the spectrum is the same as the one for isotropic
turbulence. The energy flux also becomes a constant across the
inertial range \citep{Ditlevsen11}. Some of these features will
be shown below. These and many other
properties make SABRA model a useful simplified model for 
real turbulence, which are also 
the motivation for us to choose it for our synchronisation experiments. 

We only conduct twin experiments using the SABRA model, and only
use it to investigate the relationship between the peak
wavenumber of its LLV and the threshold wavenumber for
synchronisation. The setup of these twin experiments is as 
illustrated in Fig.
\ref{fig:cases}, except that systems $A$ and $B$ are now both
the SABRA model, with different initial states. The states of the
two systems are denoted by ${\bm z}_A$ and ${\bm z}_B$. The set of shell
velocity $u_n$ with wavenumber $k_n\le k_m$ are the master modes.
The master modes in ${\bm z}_A$ are copied into ${\bm z}_B$ and replace those in 
${\bm z}_B$ at every time step in the twin experiments (c.f. Eq. (\ref{eq:ab})). 
For notational simplicity, the conditional LLE for the SABRA
model with coupling wavenumber $k_m$ is also denoted by
$\lambda(k_m)$, and it is defined by Eq.
(\ref{eq:lyadef}) with $\obu^\delta(\x, t)$ replaced by ${\bm
z}^\delta(t)$
where ${\bm z}^\delta(t)$ is the infinitesimal perturbation to
$\bm z_A(t)$. The limit of ${\bm z}^\delta(t)$ as $t\to \infty$ is the
LLV for the SABRA model.

\section{The numerical experiments \label{sect:sim}}

\subsection{Summary of the numerical experiments}
Tables \ref{tab:case1} and \ref{tab:case2} summarise the key parameters for
the LES
in Group I and II, respectively, with the definitions of the parameters given
below in subsection \ref{sect:param}. We label a case with a code that
consists of three segments. The first
segment from the left is in the form of `R$a$' with $a$ being an
integer between $1$ and $5$. Different $a$ corresponds to different kinetic
viscosity $\nu$. 
The middle segment is the acronym of the SGS stress model. The
third segment is the value of $N_{\rm LES}$, with $N_{\rm LES}^3$ being the
number of grid points of the simulation. Each case (i.e., each code) represents a
series of simulations 
with different coupling wavenumber $k_m$. Where
necessary, we append `K$b$' to the end of the code to distinguish such
simulations, with $b$ being the value of $k_m$. Also, 
the cases with codes starting with the same `R$a$' are sometimes collectively
referred to as `subgroup R$a$'.
The cases in such a subgroup may have different SGS stress models, $N_{\rm LES}$ or $k_m$. 

The parameters for the DNS  
in the three-body experiments are summarised in Table \ref{tab:dns}. They correspond to the LES in Table
\ref{tab:case1}. 
We use `R$a$DNS' with $N_{\rm DNS}$ appended to refer to the DNS in subgroup R$a$, $N_{\rm DNS}^3$ being the number
of grid points for the DNS. 
Note that each subgroup R$a$
is associated with at most a single DNS (there is no DNS for
subgroups R4 and R5). In total, we conduct about 300 simulations where two or
three velocity fields are solved concurrently (c.f. Section
\ref{sect:numeric}), with up to
$256^3$ grid points for each field.  

The parameters for the twin experiments for SABRA model are summarised in Table
\ref{tab:sabra}, with the definitions of the parameters given in Section \ref{sect:param} below. 
Ten cases with different Reynolds
numbers are computed, and they are named 
cases $Re_0$, $Re_1$, ..., $Re_{9}$, respectively.

\begin{table} % JFM
\begin{center}% JFM
\def~{\hphantom{0}}% JFM
\begin{tabular}{ccccccccccc}
$\text{Case}$ & $Re_\lambda$ & $\nu$ & $\delta t$ & $u_{\text{rms}}$ &
  $\epsilon_t$ & $\eta$ & $\lambda_a$ & $\tau_\Delta$ & $\tau_k$ & $\Delta_t$
  \\ 
\hline
R1SSM64  & 74  & 0.0060 & 0.0057 & 0.64 & 0.074 & 0.041 & 0.70 & 0.68 & 0.29 & 0.049 \\
R1SSM96  & 73  & 0.0060 & 0.0057 & 0.62 & 0.071 & 0.042 & 0.70 & 0.52 & 0.29 & 0.046 \\
R1DSM64  & 72  & 0.0060 & 0.0057 & 0.62 & 0.071 & 0.042 & 0.70 & 0.68 & 0.29 & 0.044 \\
R1DSM96  & 73  & 0.0060 & 0.0057 & 0.62 & 0.070 & 0.042 & 0.70 & 0.52 & 0.29 & 0.043 \\
R1DMM64  & 74  & 0.0060 & 0.0057 & 0.64 & 0.077 & 0.041 & 0.69 & 0.67 & 0.28 & 0.045 \\
R1DMM96  & 73  & 0.0060 & 0.0057 & 0.63 & 0.073 & 0.041 & 0.70 & 0.51 & 0.29 & 0.043 \\
R2SSM64  & 90  & 0.0044 & 0.0039 & 0.65 & 0.076 & 0.033 & 0.61 & 0.67 & 0.24 & 0.042 \\
R2SSM128 & 86  & 0.0044 & 0.0039 & 0.62 & 0.071 & 0.033 & 0.60 & 0.43 & 0.25 & 0.036 \\
R2DSM64  & 89  & 0.0044 & 0.0039 & 0.65 & 0.077 & 0.032 & 0.60 & 0.66 & 0.24 & 0.036 \\
R2DSM128 & 89  & 0.0044 & 0.0039 & 0.64 & 0.074 & 0.033 & 0.61 & 0.42 & 0.24 & 0.033 \\
R2DMM64  & 86  & 0.0044 & 0.0039 & 0.62 & 0.070 & 0.033 & 0.61 & 0.69 & 0.25 & 0.039 \\
R2DMM128 & 87  & 0.0044 & 0.0039 & 0.64 & 0.077 & 0.032 & 0.59 & 0.42 & 0.24 & 0.033 \\
R3SSM64  & 100 & 0.0030 & 0.0029 & 0.61 & 0.070 & 0.025 & 0.49 & 0.69 & 0.21 & 0.036 \\
R3SSM128 & 115 & 0.0030 & 0.0029 & 0.68 & 0.079 & 0.024 & 0.51 & 0.41 & 0.20 & 0.028 \\
R3DSM64  & 108 & 0.0030 & 0.0029 & 0.65 & 0.078 & 0.024 & 0.49 & 0.66 & 0.20 & 0.030 \\
R3DSM128 & 109 & 0.0030 & 0.0029 & 0.66 & 0.077 & 0.024 & 0.50 & 0.42 & 0.20 & 0.025 \\
R3DMM64  & 104 & 0.0030 & 0.0029 & 0.63 & 0.072 & 0.025 & 0.50 & 0.68 & 0.20 & 0.033 \\
R3DMM128 & 107 & 0.0030 & 0.0029 & 0.65 & 0.076 & 0.024 & 0.50 & 0.42 & 0.20 & 0.026 \\
\end{tabular}
  \caption{\label{tab:case1} Parameters for lower Reynolds number LESs in Group I.
  }
\end{center}
\end{table}

\begin{table} % JFM
\begin{center}% JFM
\def~{\hphantom{0}}% JFM
\begin{tabular}{ccccccccccc}
$\text{Case}$ & $Re_\lambda$ & $\nu$ & $\delta t$ & $u_{\text{rms}}$ &
  $\epsilon_t$ & $\eta$ & $\lambda_a$ & $\tau_\Delta$ & $\tau_k$ & $\Delta_t$
  \\
\hline
R4SSM32  & 401 & 0.0002 & 0.004 & 0.62 & 0.070 & 0.0033 & 0.13 & 1.10 & 0.053 & 0.045 \\
R4SSM64  & 418 & 0.0002 & 0.004 & 0.65 & 0.076 & 0.0032 & 0.13 & 0.67 & 0.051 & 0.023 \\
R4SSM128 & 459 & 0.0002 & 0.004 & 0.70 & 0.086 & 0.0031 & 0.13 & 0.40 & 0.048 & 0.012 \\
R4SSM256 & 424 & 0.0002 & 0.004 & 0.64 & 0.071 & 0.0033 & 0.13 & 0.27 & 0.053 & 0.007 \\
R4DSM32  & 420 & 0.0002 & 0.004 & 0.65 & 0.075 & 0.0032 & 0.13 & 1.08 & 0.052 & 0.037 \\
R4DSM64  & 422 & 0.0002 & 0.004 & 0.64 & 0.070 & 0.0033 & 0.13 & 0.69 & 0.053 & 0.019 \\
R4DSM128 & 450 & 0.0002 & 0.004 & 0.69 & 0.084 & 0.0031 & 0.13 & 0.40 & 0.049 & 0.010 \\
R4DSM256 & 431 & 0.0002 & 0.004 & 0.66 & 0.077 & 0.0032 & 0.13 & 0.26 & 0.051 & 0.006 \\
R4DMM32  & 443 & 0.0002 & 0.004 & 0.71 & 0.096 & 0.0030 & 0.13 & 1.00 & 0.046 & 0.049 \\
R4DMM64  & 435 & 0.0002 & 0.004 & 0.65 & 0.073 & 0.0032 & 0.13 & 0.68 & 0.052 & 0.025 \\
R4DMM128 & 414 & 0.0002 & 0.004 & 0.64 & 0.075 & 0.0032 & 0.13 & 0.42 & 0.051 & 0.013 \\
R4DMM256 & 445 & 0.0002 & 0.004 & 0.67 & 0.078 & 0.0032 & 0.13 & 0.26 & 0.051 & 0.007 \\
R5SSM32  & 569 & 0.0001 & 0.003 & 0.64 & 0.077 & 0.0019 & 0.09 & 1.07 & 0.036 & 0.045 \\
R5SSM64  & 612 & 0.0001 & 0.003 & 0.66 & 0.076 & 0.0019 & 0.09 & 0.67 & 0.036 & 0.022 \\
R5SSM128 & 663 & 0.0001 & 0.003 & 0.70 & 0.083 & 0.0019 & 0.09 & 0.41 & 0.035 & 0.011 \\
R5SSM256 & 661 & 0.0001 & 0.003 & 0.70 & 0.081 & 0.0019 & 0.09 & 0.26 & 0.035 & 0.006 \\
R5DSM32  & 627 & 0.0001 & 0.003 & 0.67 & 0.078 & 0.0019 & 0.09 & 1.07 & 0.036 & 0.037 \\
R5DSM64  & 639 & 0.0001 & 0.003 & 0.67 & 0.075 & 0.0019 & 0.10 & 0.67 & 0.037 & 0.018 \\
R5DSM128 & 648 & 0.0001 & 0.003 & 0.70 & 0.088 & 0.0018 & 0.09 & 0.40 & 0.034 & 0.009 \\
R5DSM256 & 655 & 0.0001 & 0.003 & 0.70 & 0.083 & 0.0019 & 0.09 & 0.25 & 0.035 & 0.005 \\
R5DMM32  & 570 & 0.0001 & 0.003 & 0.65 & 0.084 & 0.0019 & 0.09 & 1.04 & 0.034 & 0.050 \\
R5DMM64  & 576 & 0.0001 & 0.003 & 0.65 & 0.078 & 0.0019 & 0.09 & 0.66 & 0.036 & 0.025 \\
R5DMM128 & 572 & 0.0001 & 0.003 & 0.63 & 0.072 & 0.0019 & 0.09 & 0.43 & 0.037 & 0.012 \\
R5DMM256 & 646 & 0.0001 & 0.003 & 0.70 & 0.087 & 0.0018 & 0.09 & 0.25 & 0.034 & 0.006 \\
\end{tabular}
  \caption{\label{tab:case2} Parameters for the high Reynolds number LES
  in Group II. 
  }
\end{center}
\end{table}

\begin{table} % JFM
  \begin{center}% JFM
  \def~{\hphantom{0}}% JFM
  \begin{tabular}{ccccccccc}
    Case & $N_{\rm DNS}$ & $Re_\lambda$ & $u_{\rm rms}$ & $\epsilon$ & $\eta$ & $\nu$ & $\lambda_a$ & $\tau_k$ \\
    \hline
    R1DNS128 & 128 & 75 & 0.63 & 0.072 & 0.042 & 0.0060 & 0.71& 0.30\\ 
    R2DNS192 & 192 & 90 & 0.65 & 0.074 & 0.033 & 0.0044 & 0.61& 0.24\\ 
    R3DNS256 & 256 & 112 & 0.66 & 0.077 & 0.024 & 0.0030 & 0.51& 0.20\\
  \end{tabular}
  \caption{\label{tab:dns} Parameters for the DNS corresponding to the LES 
  in Group I. }
\end{center} 
\end{table}

\begin{table}%[H] add [H] placement to break table across pages
\begin{center}% JFM
\def~{\hphantom{0}}% JFM
\begin{tabular}{cccccccccc}
	Case & $10^5\nu$ & $K$ & $10^3\epsilon$ &  $10^3\eta$ &
         $\tau_k$ & $u_{\text{rms}}$ & $10^5 Re$ &
         $N_\textrm{SB}$ \\
        \hline
      	  $Re_0$ & $6.104$ & 0.32 & 3.9 &  2.765 & 0.125 &  0.80 &  4.1191 & 14\\
	        $Re_1$ & $3.052$ & 0.33 & 3.9 &  1.638 & 0.088 &  0.81 &  8.3154 & 15\\
          $Re_2$ & $1.526$ & 0.30 & 3.6 &  0.997 & 0.065 &  0.77 &  15.938 & 16\\
          $Re_3$ & $0.763$ & 0.31 & 3.7 &  0.587 & 0.045 &  0.78 &  32.229 & 16\\
          $Re_4$ & $0.380$ & 0.33 & 4.0 &  0.343 & 0.031 &  0.81 &  66.870 & 17\\
          $Re_5$ & $0.190$ & 0.33 & 4.0 &  0.204 & 0.022 &  0.81 &  133.99 & 18\\
          $Re_6$ & $0.100$ & 0.30 & 3.6 &  0.129 & 0.017 &  0.78 &  243.53 & 19\\
          $Re_7$ & $0.050$ & 0.32 & 3.9 &  0.075 & 0.011 &  0.80 &  502.96 & 19\\
          $Re_8$ & $0.013$ & 0.33 & 3.8 &  0.027 & 0.006 &  0.81 &  2029.3 & 21\\
          $Re_9$ & $0.003$ & 0.33 & 3.9 &  0.009 & 0.003 &  0.81 &  8162.0 & 22\\
         \end{tabular}
       \caption{\label{tab:sabra} Parameters for the SABRA model.
    }
\end{center}
\end{table}

\subsection{Definitions of main parameters \label{sect:param}}

We now summarise briefly the main parameters of the simulations. 
We introduce the energy spectrum first. Using $\bv(\x,t)$ 
and $\hat{\bv}(\k,t)$ to represent a generic DNS or LES
velocity and its Fourier transform, respectively, the energy spectrum of
$\bv$ is denoted by $E(k)$ with
\be
E(k) = \frac{1}{2}\sum_{k \le \vert \k \vert \le k+1} \lal \hat{\bv}(\k,t) \cdot \hat{\bv}^*(\k,t)\ral 
\ee 
where  $\lal ~ \ral$ represents ensemble
average, i.e., average over space and time as well as different realisations
where applicable.  
The turbulent kinetic energy $K$ and the viscous energy dissipation rate
$\ep$ then can be calculated from $E(k)$ with 
\be
K = \int_0^\infty E(k) dk, \quad \ep = 2 \nu \int_0^\infty k^2 E(k)dk.
\ee
The above equations apply to both LES and DNS, with the
understanding that $E(k) = 0$ when $k\ge k_\Delta$ for LES. 

For DNS, $\ep$ is a key parameter as it represents the mean energy
flux across the inertial range in stationary turbulence \citep{Pope00}. 
For LES, however, $\ep$ is usually not pertinent, because the energy flux is the sum of 
$\ep$ and the SGS energy dissipation  $\Pi_\Delta \equiv
-\lal \tau_{ij} \ols_{ij}\ral $ and $\ep$ is usually 
much smaller than $\Pi_\Delta$ \citep{MeneveauKatz00, Pope00}. We
refer to the sum as the total dissipation rate, and denote it by $\ep_t$, i.e., 
\be
\ep_t = \ep + \Pi_\Delta = \ep - \lal \tau_{ij} \ols_{ij}\ral. 
\ee 
$\ep_t$ is the suitable 
parameter for the characterisation of LES velocity, and it can be 
calculated from the LES velocity with $\tau_{ij}$ given by a SGS
stress model.  
As $\ep_t = \ep$ in DNS, we can use $\ep_t$ uniformly to introduce relevant length and 
time scales and other parameters that we will use to normalise the 
conditional LLEs and the threshold wavenumbers. 

We thus define the
root-mean-square (RMS) velocity $u_{\rm rms}$, the Taylor length scale
$\lambda_a$, and the Taylor scale Reynolds number via 
\be \label{eq:p1}
u_{\rm rms} = (2K/3)^{1/2}, \quad \lambda_a = (15 \nu u^2_{\rm rms}/\ep_t)^{1/2}, 
\quad Re_\lambda = \frac{u_{\rm rms} \lambda_a }{\nu},
\ee 
and the Kolmogorov length scale $\eta$ and time scale $\tau_k$ by
\be \label{eq:p2}
\eta = (\nu^3/\ep_t)^{1/4}, \quad \tau_k = (\nu/\ep_t)^{1/2}.
\ee 
The eddy turnover time at the filter
scale $\Delta$, denoted by $\tau_\Delta$, is defined as
\be  \label{eq:taudel}
\tau_\Delta = \Delta^{2/3}\ep_t^{-1/3}.
\ee 
For DNS, the definitions introduced in Eqs. (\ref{eq:p1}) and
(\ref{eq:p2}) are the same as the usual definitions. 
For LES, $\lambda_a$, $\eta$ and $\tau_k$ 
do not usually represent real scales in the velocity field. Rather they should be understood as
approximations to those of the true velocity which the LES is
simulating. Though very often LES results are normalised by
parameters at the filter scale (e.g., $\tau_\Delta$), we also use
$\eta$ and $\tau_k$ defined above in order to
compare the results with those in the literature and to present
the results in a unified way at both low and high Reynolds
numbers. 

As will be shown later, it is instructive to correlate
the LES results (in particular the threshold wavenumber $k_c$) using an
effective dissipation length scale $\Delta_t$. We firstly define a (constant)
effective viscosity $\nu_t$ by 
\be 
\nu_t \equiv \frac{ \ep_t}{2 \lal \ols_{ij} \ols_{ij}\ral}.
\ee
Since the above equation implies
\be  \label{eq:eff}
\ep_t = 2 \nu_t \lal \ols_{ij}\ols_{ij}\ral,
\ee 
it is clear that $\nu_t$ is defined in such way that the total
energy dissipation can be calculated from the eddy viscosity model $\tau_{ij}= -2 \nu_t \ols_{ij}$. 
The effective dissipation length scale $\Delta_t$
is then defined as 
\be \label{eq:delta_t}
\Delta_t = (\nu_t^3/\ep_t)^{1/4}.
\ee 
According to the Kolmogorov phenomenology, $\Delta_t$ should scale with
$\Delta$ at high Reynolds numbers for $\Delta$ in the inertial range. 
Nevertheless, our numerical experiments later
show that $\Delta_t$ correlates better with data than $\Delta$. 

For the SABRA model, similar quantities can be defined. 
The turbulent kinetic energy $K$ is defined by \citep{Biferale03, Ditlevsen11}
\be
K = \frac{1}{2} \sum_{n=1}^{N_\textrm{SB}} \lal \vert u_n \vert^2 \ral,
\ee
whereas the viscous energy dissipation rate $\ep$ is defined by
\be
\ep = \nu \sum_{n=1}^{N_\textrm{SB}} k_n^2\lal \vert u_n \vert^2\ral. 
\ee
We then define $u_\textrm{rms}$ using Eq. (\ref{eq:p1}), and $\eta$ and $\tau_k$ using Eq. (\ref{eq:p2}), with
the understanding that $\ep_t = \ep$ for the SABRA model. 
The Taylor scale Reynolds number is not often
used in SABRA model. Instead, we introduce a crude `integral'
length scale $L \equiv \pi/k_1 \approx 31.42$, where $k_1$ is the wavenumber for
the first shell, and define the Reynolds number as
\be
Re = \frac{u_\textrm{rms} L}{\nu}. 
\ee
The values of these parameters for the LES are summarised in
Tables \ref{tab:case1} and \ref{tab:case2}, those for DNS in
Table \ref{tab:dns}, and the SABRA model Table \ref{tab:sabra}. 

\subsection{Numerical methods \label{sect:numeric}}

Information for the numerical schemes is given in what follows. The NSE and
the fNSE (Eqs. (\ref{eq:nse}) and (\ref{eq:fnse})), together with
the continuity equations,
are integrated with the pseudo-spectral method. 
The domain is discretised uniformly with $N^3_{\textrm{DNS}}$ and
$N^3_{\textrm{LES}}$ 
grid points for DNS and LES, respectively.
The two-thirds rule \citep{Pope00} is applied to de-alias 
the advection term so that the maximum
effective wavenumber is $N_{\textrm{DNS}}/3$ for DNS and
$N_{\textrm{LES}}/3$
for LES. 
Time stepping uses an explicit second-order Euler scheme \citep{Lietal20}. The
viscous diffusion term is treated with an integration factor
prior to
time discretization. The step-size $\delta t$ is chosen in such
a way that the Courant-Friedrichs-Lewy number $ u_{\rm rms} \delta
t/\delta x$ is less than $0.1$, where $\delta x$ is the grid size. 

The conditional LLEs are calculated using the algorithm
reported in, e.g., \citet{BoffettaMusacchio17, BudanurKantz22},
with some minor modification. We
run concurrent simulations for $\obu_{A}$ and
$\obu_{B}$ in twin experiments, and for $\u_M$, $\obu_A$ and $\obu_B$
in three-body experiments. $\obu_A$ and $\obu_B$ initially differ by a small
perturbation such that 
\be
e \equiv \Vert \obu_A(\x,0) - \obu_B(\x,0)\Vert  
\ee
is small, and the perturbation is only introduced to the slave
modes in system $B$. 
During the simulation, $\obu_{B}$ is periodically
re-initialised at $t_n = n \Delta t$ ($n=1,2,...$) for some short time
interval $\Delta t ( \ne \delta t)$ so that
\be \label{eq:rescale}
\obu_{B}(\bx, t_n^+) = \obu_{A}(\bx, t_n) + g_n^{-1} \left[\obu_{B}(\bx,
t_n^-) - \obu_{A}(\bx, t_n)\right], 
\ee 
where $t_n^\pm$ are the times immediately after/before $t_n$ and 
\be 
g_n = e^{-1}\Vert \obu_{B}(\bx, t_n^-)- \obu_{A}(\bx, t_n)\Vert
\ee 
is the amplification factor for the perturbation. The conditional LLE is given by
\be 
\lambda (k_m) \approx \lim_{M\to \infty} \frac{1}{M \Delta t} \sum_{n=1}^M
\log g_n. 
\ee 
In our simulations, $\Delta t \approx 0.1 \tau_k$ for cases in group I. It
varies somewhat for cases in group II but is never larger than
$0.05 \tau_\Delta$.

Note that, because of the coupling between $\obu_A$ and $\obu_B$ (c.f., Eqs.
(\ref{eq:abm}) and (\ref{eq:ab})), the perturbation $\obu_{B}-\obu_{A}$ is
always zero for Fourier modes with wavenumber $\vert \k\vert \le k_m$, as 
required by Eq.
(\ref{eq:cle0}). 
The LLE for filtered DNS, namely $\lambda_\Delta$, is calculated
in much the same way. Here we run two uncoupled DNS for $\bu_A$ and
$\bu_B$. When $\bu_B$ is re-initialised, the perturbation
in Fourier modes with $\vert \k \vert > k_\Delta$ is set to zero by letting
\be
\hat{\bu}_B(\k, t) = \hat{\bu}_A(\k,t) \quad (\vert \k \vert > k_\Delta), 
\ee
hence enforcing Eq. (\ref{eq:filpert}).

The SABRA model is integrated in time with an explicit fourth order Runge-Kutta method with
step-size $\delta t_\textrm{SB} = 10^{-5}$. 
The conditional LLEs for the SABRA model are calculated using the
same algorithm described above, with $\obu_A$ and $\obu_B$
replaced by respective copies of the SABRA model. 

\section{Results and analyses \label{sect:rd}}

\bfig
\centering
\ig[width=0.48\lnw]{Ek_R3} %
\ig[width=0.48\lnw]{Ek_R5}
\caption{\label{fig:Ek} The energy spectra. Left: For cases in R3. Right: For
cases in R5. Dash-dotted lines without symbols: the $k^{-5/3}$ power law.
Dashed lines without symbol: DNS for R3.} 
\efig

We begin this section with a preamble to present
several supplementary results. 
Firstly, we present the energy spectra for 
selected cases as a validation of the numerics. Fig. \ref{fig:Ek} shows
the energy spectra, with lower Reynolds number results 
on the left panel and high Reynolds number results on the
right panel. The DNS spectrum included in the left panel 
displays features that are consistent with
the $k^{-5/3}$ Kolmogorov spectrum. The results from LES
reproduce the main features of the DNS spectrum. DSM and DMM in
particular appears to capture the tail of the
spectrum better. No DNS data are available for comparison in 
the high Reynolds number cases, but 
an extended range with a slope of $-5/3$ is clearly
visible in the spectra, reproducing the expected 
behaviours from these canonical SGS stress models.

As explained previously, several questions auxiliary to our main objective 
are addressed in this investigation. 
One of them is how different coupling methods impact the
synchronisation process. Specifically, one may ask whether the 
synchronisation threshold is independent of the coupling
mechanism, or more pointedly, whether synchronisation in three-body
experiments can be described by the conditional LLEs of the slave
system at all, noting that the actual measure of
synchronisation is the synchronisation error, not the conditional
LLE. To keep the 
focus of the paper clear, we present 
the discussion of these questions in the appendices, 
and only relate the conclusions as follows:
\begin{enumerate}
\item 
Firstly, the
conditional LLEs of the slave system still characterise the
synchronisation process in three-body experiments. Specifically,
the decay rate of the synchronisation error is still
approximately equal to 
the conditional LLE.
\item 
Secondly, the threshold wavenumbers 
for the three-body and the twin experiments are essentially the
same. Therefore, the synchronisation process is insensitive to
the nature of the master flow. 
\end{enumerate}

The above conclusions show that we may bypass the results on 
synchronisation error and focus on the conditional LLEs. Also, we
do not need to distinguish the three-body from the twin
experiments as far as the synchronisation threshold and the
conditional LLEs are concerned. We will follow these principles in
what follows.

\subsection{The relationship between the unconditional LLEs for LES and filtered
DNS \label{sect:ule}}

This subsection is devoted to the unconditional
LLE for LES, i.e., $\lambda(k_m=0)$, and  
its relationship with the LLE for filtered DNS,
$\lambda_\Delta$. 
To the best of our knowledge, the LLE for filtered DNS has not
been reported before. 
Only unconditional LLE is considered in this subsection so 
we will simply use $\lambda$ to represent $\lambda(k_m = 0)$.

\bfig
\centering
\ig[width=0.48\lnw]{lya_tauk_fDNS_group1} %
\ig[width=0.48\lnw]{lya_tauk_fDNS_group2}
\caption{\label{fig:lya_tauk_fDNS} Normalised unconditional LLEs $\lambda
\tau_k$ as functions of $\eta/\Delta$ for the cases in Group I (the left panel) 
and Group II (the right panel). The horizontal dash-dotted line indicates
$\lambda \tau_k = 0.12$ corresponding to
the value for DNS with moderate Reynolds number
\citep{BoffettaMusacchio17, BereraHo18, Inubushietal2023}.} 
\efig

\bfig
\centering
\ig[width=0.5\lnw]{PDT_highRe_fitting}
\caption{\label{fig:PDT_highRe_fitting} Normalised unconditional LLE 
$\lambda \tau_\Delta$ as functions of $\eta/\Delta$ for cases in
subgroups R4 and R5. Symbols: simulations results. Lines: least squares fitting based on Eq.
\ref{eq:scaling1}. }
\efig

The main result is shown in Fig. \ref{fig:lya_tauk_fDNS}, which
compares the LLEs of LES with those of filtered DNS, with 
the cases in Group I plotted in the 
left panel and those in group II in the right panel. 
The normalisation parameters are given in Tables \ref{tab:case1} --
\ref{tab:dns}.
In order to compare with the results  
shown in Fig. 1 of \citet{BudanurKantz22} and Fig. 7 of
\citet{Nastacetal17}, we use $\tau_k$ to normalise $\lambda$, and 
plot $\lambda \tau_k$ against $\eta/\Delta$. 
There is no DNS data to compare the high Reynolds number LES
results with. We plot the same set of low Reynolds number DNS results 
in the right plane, merely to highlight the scaling relation.  

The most salient feature in the figure is the power law behaviour
emerging for smaller $\eta/\Delta$. For the lower Reynolds number
data, the scaling law can be observed for $\eta/\Delta$
being less than a value between $0.1$ and $0.2$ for the three SGS stress models. 
For filtered DNS, it is observed for $\eta/\Delta \lesssim 
0.08$. For the high Reynolds number data, the scaling
behaviour appears to be present for all $\eta/\Delta$ values
covered in our simulations. 

No scaling law can be discerned for larger $\eta/\Delta$ values
for either LES or filtered DNS results. It appears that this
range of values for $\eta/\Delta$ coincide
with those in \citet{Nastacetal17} and
\citet{BudanurKantz22}, which might explain why no
scaling regime was shown in their results. 
In this range, the LLE for the LES generally increases with
$\eta/\Delta$. This trend is consistent with 
\citet{Nastacetal17} and \citet{BudanurKantz22}. In addition, our
results also show
that the LLE is different for different models, with SSM and DMM
producing the smallest and largest values, respectively. In
contrast to the LLEs for LES, the LLE of the filtered DNS 
actually remains roughly a constant, a trend that is captured
better by the DMM. 

Returning back to the power law region, the data suggest that
the three models display the same scaling exponent, and the
scaling exponent is close to that of the filtered DNS. This
observation can be 
quantified by fitting a power-law in the form of  
\be \label{eq:fit1}
 \lambda \tau_k = a (\eta/\Delta)^b 
\ee
to the data. The parameters $a$ and $b$ are found by least
squares fitting applied to the transformed data $(\ln \eta/\Delta, \ln
\lambda\tau_k)$. The values for $a$ and $b$ and Eq.
\ref{eq:fit1} are also shown in Fig. \ref{fig:lya_tauk_fDNS}. 
Eq. \ref{eq:fit1} appears as a straight line with slope $b$ in
the figure.
The data in the left panel shows that the slopes for the LES results
are slightly steeper than the one for filtered DNS. On the other
hand, the slopes
for the LES results at higher Reynolds numbers, shown in the
right panel, are slightly smaller, though we should note
this is compared to data for filtered DNS 
obtained at lower Reynolds numbers. In any case, it is justified to
say that the differences in the slopes between different cases are small. 

There are conspicuous differences for 
coefficient $a$ between SGS models and the filtered
DNS. However, given the close agreement for the scaling exponent
$b$, 
we argue that the LLE for an LES should be interpreted as
an approximation to the LLE for the filtered DNS that the LES is
simulating.
This relationship, as intuitive as it may seem, has
not been reported before. 
In this view, the LLE for an LES is a measure of the growth
rate of the infinitesimal perturbation confined in the scales
larger than $\Delta$ in the DNS.

The above claim can be supported by the following
phenomenological analysis.  
In fact, Kolmogorov phenomenology
suggests that
\be
\lambda \sim 
\tau_\Delta^{-1}, \qquad \lambda_\Delta \sim \tau_\Delta^{-1},
\ee
so that $\lambda \tau_\Delta \sim \text{constant}$ for $\Delta$ in the inertial
range. In light of Eq. \ref{eq:taudel}, this would imply $\lambda \tau_k \sim
\Delta^{-2/3} \sim (\eta/\Delta)^{2/3}$. Therefore, according to
the Kolmogorov phenomenology, $\lambda\tau_k$ should decrease as $\Delta$ increases, and the slope for $\lambda \tau_k$
should be $2/3$. The empirical values of
the exponent $b$ again are close to this semi-analytical value.
This phenomenological argument applies equally to $\lambda$ and
$\lambda_\Delta$, which shows that there is a physical basis
for the two to be same. 

\begin{table} % JFM
  \begin{center}% JFM
  \def~{\hphantom{0}}% JFM
  \begin{tabular}{cccc}
  Coefficients & SSM & DSM & DMM\\
\hline
  $(\alpha, \beta, c_1)$ & $(0.05,0.21,0.12)$ & $(0.05, 0.10,0.29)$ & $(0.08,0.16,0.29)$
  \end{tabular}
    \caption{\label{tab:exp} Coefficients for the fitting function of the data in Fig.
    \ref{fig:PDT_highRe_fitting} (c.f., Eq. (\ref{eq:scaling1})).
    }
  \end{center}
  \end{table}
  
There is non-trivial discrepancy between the empirical values for
$b$ and $2/3$ (observed for both filtered DNS and LES results). We explore this issue further with 
the high Reynolds number simulations in
Group II, and demonstrate below that the deviation can be
plausibly explained with current theories. 
We start by plotting the results for $\lambda
\tau_\Delta$, which are shown with the symbols in Fig. \ref{fig:PDT_highRe_fitting}. 
To quantify the deviation, we note
Eqs. \ref{eq:taudel} and
(\ref{eq:fit1}) imply
\be \label{eq:lmdel}
\lambda \tau_\Delta = a \left(\frac{\eta}{\Delta}\right)^{\alpha}, 
\quad \alpha = b - \frac{2}{3}. 
\ee
Thus $\alpha$ can be found from the values of $b$ we obtained previously (c.f., the right panel of 
Fig. \ref{fig:lya_tauk_fDNS}). $\alpha$ obviously is smaller than
$b$, indicating weaker dependence on $\eta/\Delta$. This is
consistent with Fig. \ref{fig:PDT_highRe_fitting}. 
Fig. \ref{fig:PDT_highRe_fitting} also shows a systematic difference 
between results for subgroups R4 and R5 which have different
Reynolds numbers (which thus have been plotted 
separately). 
This observation suggests that $a$ 
in Eq. (\ref{eq:lmdel}) depends on the Reynolds number, 
and we may write empirically
\be \label{eq:scaling1}
\lambda \tau_\Delta = c_1 \left(\frac{\eta}{\Delta}\right)^{\alpha} Re_\lambda^\beta,
\ee 
for some constants $c_1$ and $\beta$. The values for $c_1$ and $\beta$ are found with 
linear regression, and are given in Table \ref{tab:exp}. 
Eq. (\ref{eq:scaling1}) with coefficients given in Table \ref{tab:exp} is 
shown with lines in Fig. \ref{fig:PDT_highRe_fitting}. 

To compare the above scaling relation with results in the literature, we 
rewrite Eq. (\ref{eq:scaling1}) in terms of the Reynolds number
defined with an integral 
length scale $L$. $L$ can be defined via the energy spectrum as
in, e.g., \citet{Batchelor53}, though its precise formula is not important
for the scaling argument
below. Introducing the dissipation
constant $C_\ep \equiv L \ep_t/ u_{\rm rms}^3$,  
it is easy to show that Eq. (\ref{eq:scaling1}) implies
\be 
\lambda \tau_\Delta 
\sim C_\ep^{-\beta/2-\alpha/4} Re_L^{\beta/2 - 3\alpha/4} \left(\frac{\Delta}{L}\right)^{-\alpha},
\ee
where $Re_L \equiv u_{\rm rms} L/\nu$ is the integral scale
Reynolds number. 
As is reviewed in \citet{Vassilicos15}, 
the dissipation constant $C_\ep$ decreases with increasing $Re_L$
though perhaps
only weakly at high Reynolds numbers. It also displays 
interesting scaling behaviours reported in
non-equilibrium turbulence \citep{ValenteVassilicos12}.  
With the values of $\alpha$ and $\beta$ obtained  
above, we then find 
\be \label{eq:lmdel_final}
\lambda\tau_\Delta \sim 
\begin{cases}
  ~~ C_\ep^{-0.12} Re_L^{0.06}\displaystyle \left(\frac{\Delta}{L}\right)^{-0.05} & \text{for SSM,}\\
  ~~ C_\ep^{-0.06} Re_L^{0.01}\displaystyle \left(\frac{\Delta}{L}\right)^{-0.05} & \text{for DSM,} \\
  ~~ C_\ep^{-0.10} Re_L^{0.02}\displaystyle \left(\frac{\Delta}{L}\right)^{-0.08}  & \text{for DMM.}
\end{cases}
\ee
Note that $C_\ep$ tends to decrease as $Re_L$ increases. Therefore, $\lambda \tau_\Delta$ 
increases with the Reynolds number and increases when $\Delta/L$ is
decreased. Both are corrections to the Kolmogorov scaling. As one may see,
they are both relatively small, not unexpected. 

The corrections are reminiscent of what is observed for the Lyapunov exponents of
unfiltered DNS \citep{BoffettaMusacchio17,BereraHo18}, where
small deviation from Kolmogorov scaling is also found. 
The quantity corresponding to $\lambda \tau_\Delta$ in those investigations is
$\lambda_{\rm DNS}\tau_k$, where $\lambda_{\rm DNS}$ denotes the LLE for the
unfiltered DNS. It is found 
that $\lambda_{\rm DNS} \tau_k \sim Re_L^{0.14}$ 
\citep{BoffettaMusacchio17,BereraHo18,
Inubushietal2023}, whereas Kolmogorov phenomenology
suggests $\lambda_{\rm DNS} \tau_k \sim $ constant.  

The origin of the deviation is debated. 
One might attribute it to small-scale intermittency (for reviews of small scale
intermittency in turbulence, see, e.g. \citet{Frisch95,SreenivasanAntonia97}). 
One might reasonably expect the intermittency correction to increase with $Re_L$ which thus might
account for the $Re_L$ factor in Eq. \ref{eq:lmdel_final}. 
Meanwhile, intermittency is stronger for smaller filter scale $\Delta$. 
This is also consistent 
with Eq. (\ref{eq:lmdel_final}). However, it should be noted that the multi-fractal formalism predicts opposite dependence on $Re_L$ \citep{BoffettaMusacchio17}.

Another possible origin of the deviation is the
sweeping effect, as argued recently in \citet{Geetal23}. 
The sweeping effect introduces 
the sweeping time scale $\eta/u_{\rm rms}$ (also known as the Eulerian time scale) 
into the parametrisation of $\lambda_{\rm DNS}\tau_k$.  
The same argument, applied to $\lambda \tau_\Delta$, suggests
that we should introduce a sweeping time scale 
$\tau^E_\Delta = \Delta/u_{\rm rms}$ into the
parametrisation of $\lambda\tau_\Delta$, which thus implies
$\lambda \tau_\Delta$ is a function of
$\tau^E_\Delta/\tau_\Delta$. As a
demonstration, let us assume $\lambda
\tau_\Delta \sim (\tau^E_\Delta/ \tau_\Delta)^c$ for some
constant $c$. We then obtain
\be
\lambda\tau_\Delta \sim  C_\ep^{c/3}
\left(\frac{\Delta}{L}\right)^{c/3}.
\ee
As
$C_\ep$ decreases with $Re_L$ \citep{GotoVassilicos09, Vassilicos15}, a negative
$c$ can potentially provide an explanation
for the behaviour of $\lambda \tau_\Delta$ expressed in Eq.
(\ref{eq:lmdel_final}).  

Finally, the dependence of $\lambda
\tau_\Delta$ on $Re_L$ can also be a finite
Reynolds number effect. If the ratio 
$\tau_k/\tau_\Delta$ is not negligible at our Reynolds numbers,
$\lambda \tau_\Delta$ should be a function of
$\tau_k/\tau_\Delta$, hence depends on $Re_L$ and $\Delta/L$. 
To sum up, there are 
competing theories for the deviation from Kolmogorov phenomenology 
observed in $\lambda \tau_\Delta$. However, 
the fact that the deviation can be explained
plausibly by these
theories lends significant support to the contention that 
the scaling 
behaviour observed in 
$\lambda$ is not incidental, but rather 
originates from the fact that $\lambda$ is 
an approximation to $\lambda_\Delta$. 

The knowledge about the relationship between $\lambda$ and
$\lambda_\Delta$ can be beneficial. 
On this regard, one is tempted to compare $\lambda$ (or
$\lambda_\Delta$)
with the finite size Lyapunov exponents (FSLEs)
\citep{Boffettaetal02a, BoffettaMusacchio17}. The FSLE
measures the growth rate of a finite size velocity perturbation $\delta v$ by
calculating the doubling time of the perturbation. It
is found that the FSLE is $\sim (\delta v)^{-2}$ in isotropic turbulence for
$\delta v$ corresponding to fluctuations over length scales in the inertial
range. There is significant difference between the LLEs of an LES velocity
field and the FSLE, as the LLEs
measures the growth of infinitesimal perturbation. Nevertheless, if we
assume that the most unstable velocity perturbation scales with $\Delta$ according
to Kolmogorov scaling, we have $\Vert \obu^\delta \Vert \sim \Delta^{1/3}$.
Ignoring the dependence on $Re_L$ and $\Delta/L$ shown in Eq. (\ref{eq:lmdel_final}), we 
have $\lambda \sim \tau_\Delta^{-1}$,
thus 
\be
\lambda \sim \tau_\Delta ^{-1} \sim \frac{\Vert \obu^\delta \Vert}{\Delta}
\sim \Vert \obu^\delta \Vert ^{-2}, 
\ee
which mimics the scaling law for the FSLEs. This 
argument hints at the possibility of using LES 
to gain insight into the FSLEs.  

\subsection{Conditional LLEs and the threshold 
coupling wavenumber \label{sect:cle}}

We now move on to the results for the conditional LLEs $\lambda(k_m)$. 
Two flows coupled with coupling wavenumber $k_m$ can synchronised only when
$\lambda(k_m) < 0 $. One main parameter of interest 
is the threshold value $k_c$, i.e., the $k_m$ value for
which $\lambda(k_m) = 0$. We will present the results for
$\lambda(k_m)$ first, and then discuss those for $k_c$. 

 \bfig
	\centering
	\ig[width=0.48\lnw]{lya_km_deltat_R1} %
	\ig[width=0.48\lnw]{lya_km_deltat_R3}
	\ig[width=0.48\lnw]{lya_km_deltat_R4} %
	\ig[width=0.48\lnw]{lya_km_deltat_R5}
	\caption{\label{fig:lya_km_deltat} Normalised conditional LLEs
  $\lambda \tau_\Delta$ as functions of $k_m \Delta_t$. Top-left: R1. 
  Top-right: R3. Bottom-left: R4. Bottom-right: R5. }
\efig

The data for $\lambda (k_m) \tau_\Delta$ 
are plotted against $k_m \Delta_t$ in Fig. \ref{fig:lya_km_deltat}, where $\Delta_t$ is 
the effective dissipation length defined in Eq. (\ref{eq:delta_t}). 
The reason why $\Delta_t$ is used to normalise $k_m$
will become clear when we discuss the results for $k_c$. 
Only the data from subgroups R1, R3, R4, and R5 are shown in
Fig. \ref{fig:lya_km_deltat}. The results for R2 have been omitted as 
they fall somewhere between those of R1 and R3.

In all cases $\lambda(k_m) \tau_\Delta$ is a decreasing function of $k_m \Delta_t$. 
For the low Reynolds number cases in R1 (top-left) and R3 (top-right), 
the results for SSM and DSM only change slightly with $Re_\lambda$ or 
the filter scale $\Delta$. In contrast, those for DMM 
depend on the two parameters quite significantly. 
For DMM, $\lambda \tau_\Delta$ is larger for larger $\Delta$, and this effect 
is the strongest for cases in R3 which have larger Reynolds numbers. 
Data for high Reynolds number cases are shown in the two bottom panels of
Fig. \ref{fig:lya_km_deltat}. 
The results depend on $Re_\lambda$ and $\Delta$ to some extent, 
but the variation is moderate, and it is reduced as $k_m\Delta_t$ increases. 
The curves collapse well as $\lambda(k_m)\tau_\Delta$ approaches zero. 

\bfig
\centering
\ig[width=0.6\lnw]{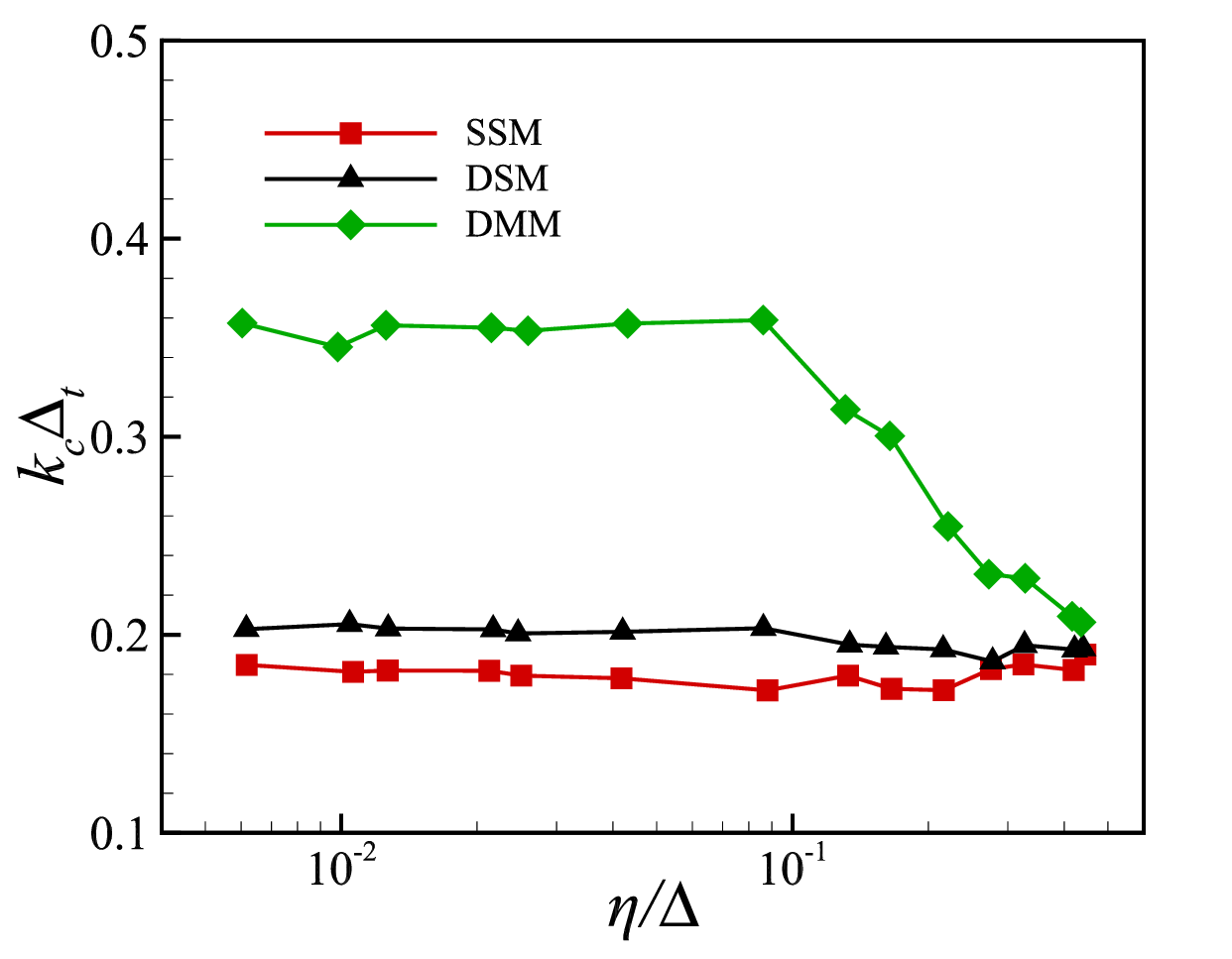} %
\caption{\label{fig:kcetat} The threshold wavenumber $k_c \Delta_t$ for all subgroups R1 to R5. }
\efig

\bfig
\centering
\ig[width=0.48\lnw]{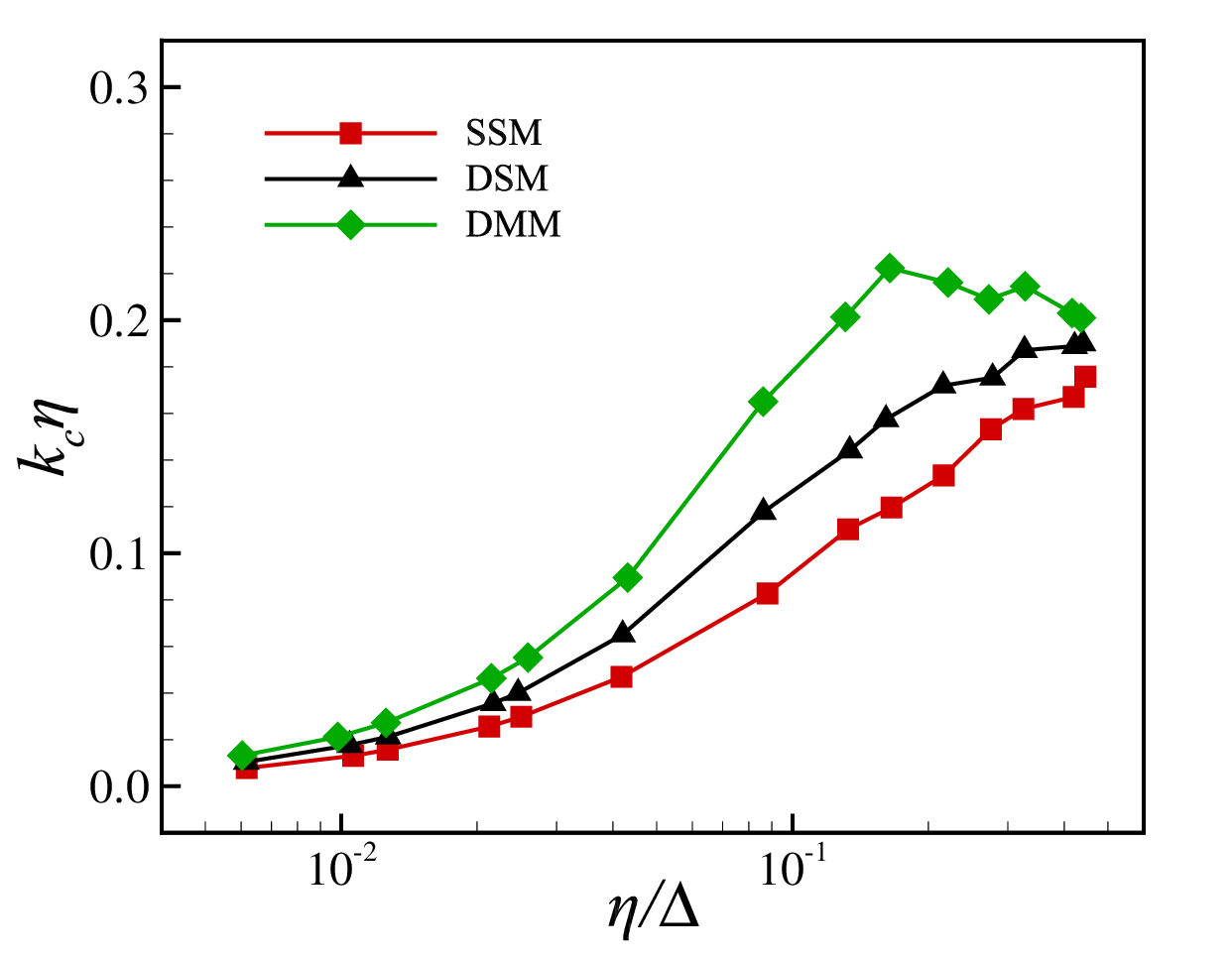}
\ig[width=0.48\lnw]{Dleta_Dletat} 
\caption{\label{fig:kceta} Left: The threshold wavenumber $k_c \eta$ for all subgroups R1 to R5. Right:
$\Delta_t/ \Delta$. The slope of the dashed line is $1$. }
\efig
 
The dimensionless threshold wavenumber $k_c \Delta_t$ is the
value of $k_m\Delta_t$ where the curve for $\lambda(k_m)$ crosses
the horizontal axis. The values of $k_c\Delta_t$ are read from Fig.
\ref{fig:lya_km_deltat}, 
and plotted as a function of $\eta/\Delta$ in 
Fig. \ref{fig:kcetat}. 
The interesting observation is that 
$k_c \Delta_t$ for SSM and DSM are almost constant across 
the whole range of values for $\eta/\Delta$. 
For SSM, $k_c\Delta_t \approx 0.18$, and 
$k_c \Delta_t\approx 0.20$ for DSM. These values, incidentally, are close to the DNS
value of $k_c \eta \approx 0.2$. 
For DMM, $k_c\Delta_t$ remains approximately a constant $0.36$ for 
$\eta/\Delta$ up to approximately $0.08$, and decreases towards
$0.2$ as $\eta/\Delta$ increases further. 
Therefore, for large $\eta/\Delta$, $k_c \Delta_t$ approaches the DNS 
value $0.2$ for all models, 
which is to be expected since LES approaches DNS when $\Delta \to
\eta$. 
At a given $\eta/\Delta$, $k_c \Delta_t$ is the
smallest for SSM, somewhat larger for DSM, and the largest for DMM. 

We may contrast the results for $k_c \Delta_t$ with those for
$k_c \eta$. The values of $k_c\eta$ 
are plotted against $\eta/\Delta$ in the left panel of Fig. \ref{fig:kceta}. 
The main observation is $k_c\eta$ increases monotonically with $\eta/\Delta$ for SSM and
DSM, while for DMM, it increases monotonically up to a value close to $0.2$ at
$\eta/\Delta \approx 0.13$, and then undulate for larger
$\eta/\Delta$. 

The normalisation $k_c \eta$ has the advantage of being more
intuitive, as the behaviour of $\eta$ is well-known. However,
$k_c\eta$ approaches zero for all models as
$\eta/\Delta \to 0$, which obscures the differences between the
models at small $\eta/\Delta$. On the other hand, $k_c \Delta_t$
is able to capture the differences, and it displays better
asymptotic behaviours as $\eta/\Delta \to 0$. 
Based on Fig. \ref{fig:kcetat}, we can extrapolate the results to
much smaller $\eta/\Delta$, but it is difficult to do so based on Fig.
\ref{fig:kceta}. 

To gain more insight into the above results, it is of interest
to look into the behaviours of $\Delta_t$. $\Delta_t/\Delta$ 
is shown in the right panel of Fig. \ref{fig:kceta} as a function
of $\eta/\Delta$. We can
see $\Delta_t/\Delta$ approaches constants as $\eta/\Delta \to 0$. When
$\eta/\Delta \to 1$, the curves tend towards $\Delta_t/\Delta = \eta/\Delta$,
as can be seen from the dashed line.
Therefore, roughly speaking, $\Delta_t$ is an interpolation between
$\eta$ and $\Delta$. 
The asymptotic values for $\Delta_t/\Delta$ as $\eta/\Delta \to 0$ can be read
from the figure. Using these values, we can find that, as
$\eta/\Delta \to 0$, $k_c\Delta$ approaches approximately $0.16$, $0.13$,
and $0.22$ for SSM, DSM, and DMM, respectively.     

The data for $k_c \Delta_t$ obtained in this subsection will be
used to examine their relationship with the peak wavenumbers of
the LLVs in the next subsection, which is our main objective. These data by
themselves mainly serve to reveal the difference in the 
synchronisability of DNS and LES. The data could also supplement
previous discussions related to the predictability of LES
and the dynamic contents in LES \citep{Nastacetal17, BudanurKantz22},
which are important questions from the perspective of turbulent
modelling. 

\subsection{Threshold wavenumbers and the leading Lyapunov vectors \label{sect:lv}}

\bfig
\centering
\ig[width=0.6\lnw]{LV_DNS_k0_keta}
\caption{\label{fig:LV_DNS} The average energy spectra of the unconditional
LLV from the DNS data in subgroups R1, R2 and R3. 
The spectra are normalised so that the total energy is
unity. The vertical line marks the value $k\eta = 0.2$. }
\efig

As is reviewed in the Introduction, evidence 
suggests that the synchronisation threshold wavenumber obtained 
in the previous subsection should be
the same as the peak wavenumber for the energy spectrum of the 
unconditional 
LLV for the slave system. 
Though the amount of evidence is significant, 
in this subsection, we
further examine this relationship using LES synchronisation 
experiments
(and later with the SABRA model). As we explain in the
Introduction, here LES is treated as a chaotic dynamical system
that is different from DNS. Our thesis is that the
relationship can be substantiated by showing it can also be
observed in systems different from DNS. For
brevity, we use LLVs to refer to the unconditional LLVs.

We denote the average energy
spectrum of the LLV by
$E_\Delta(k)$.
To give a first impression of the relationship, we plot
in Fig. \ref{fig:LV_DNS}
$E_\Delta(k)$ from our DNS data. 
This figure is
essentially the same as Fig. 2(d) of \citet{BudanurKantz22}, 
although here we use a
linear scale for $E_\Delta(k)$ to highlight the profiles around the
peaks of the spectra. 
The figure shows that the
spectra do peak approximately at $k\eta \approx 0.2$, which
coincides with
the threshold wavenumber found in synchronisation experiments for
isotropic turbulence. 

\bfig
\centering
\ig[width=0.45\lnw]{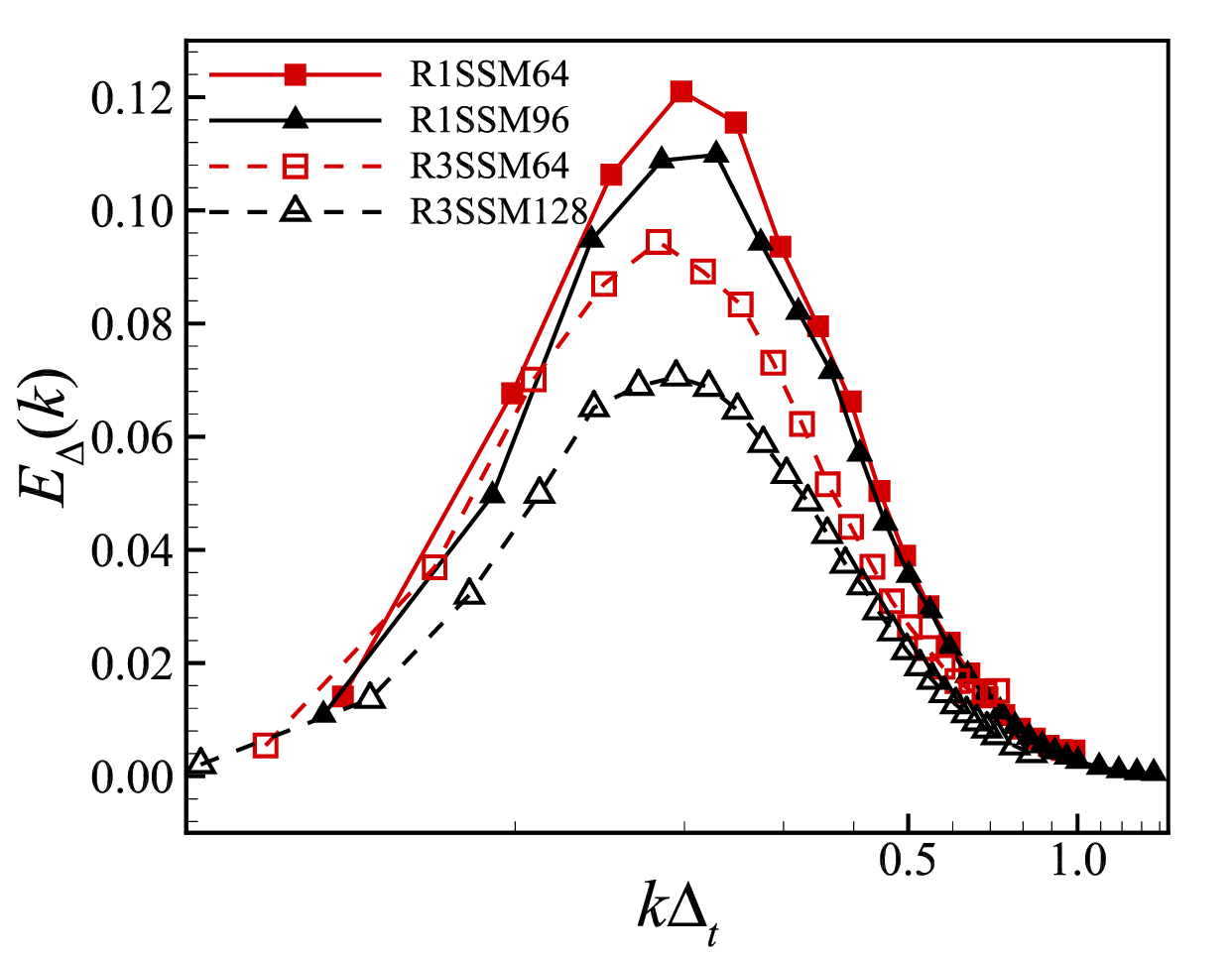}%
\ig[width=0.45\lnw]{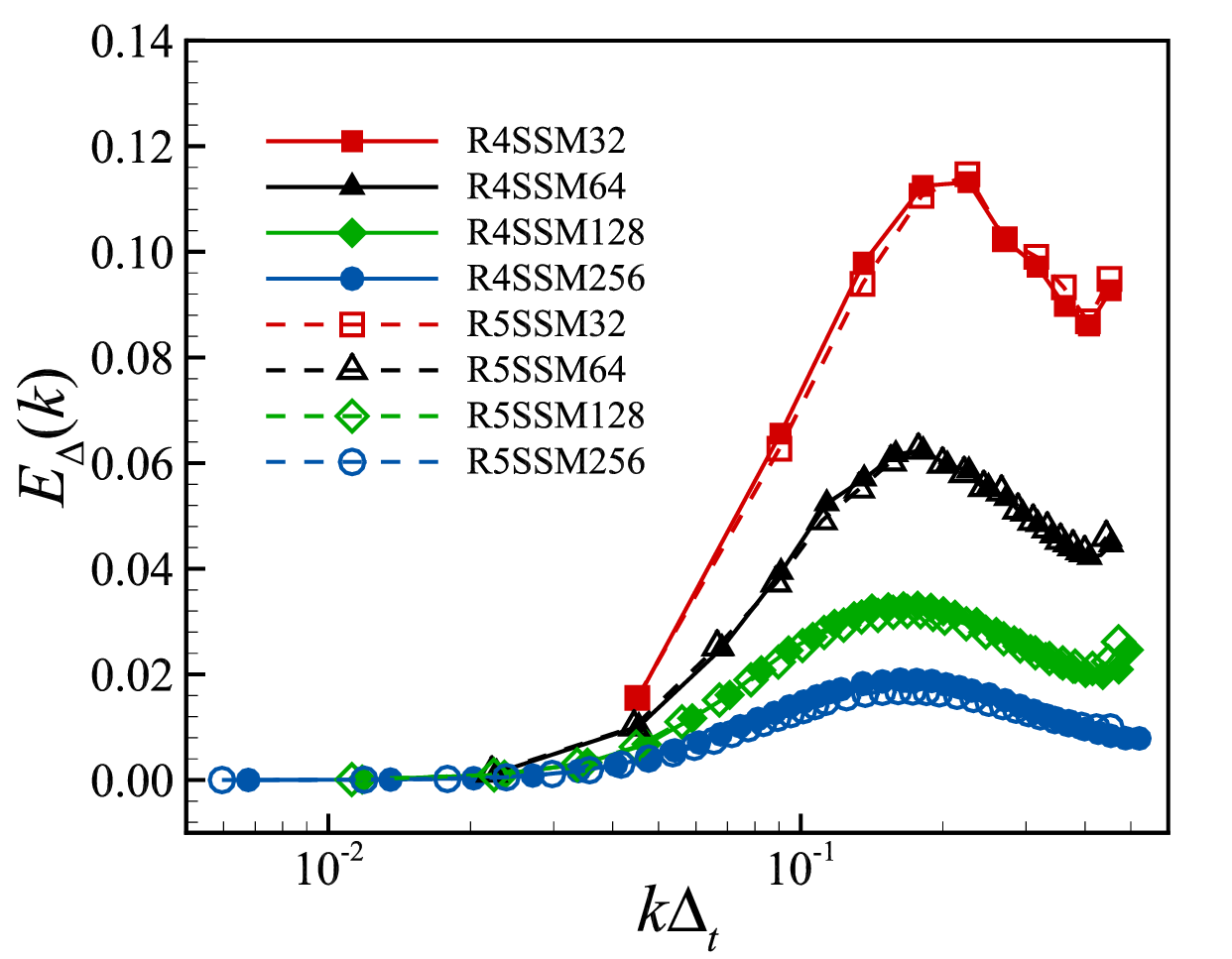}
\ig[width=0.45\lnw]{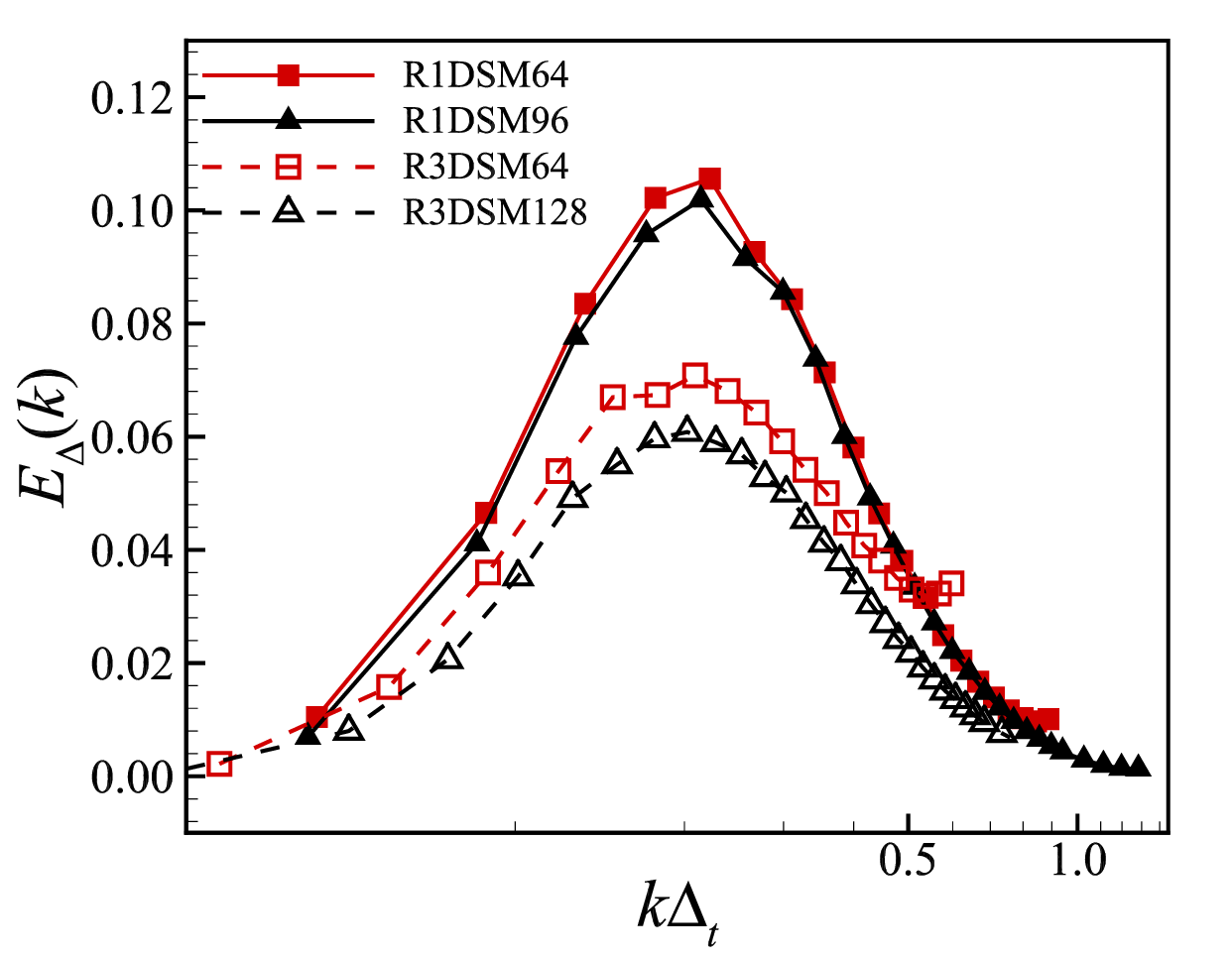}%
\ig[width=0.45\lnw]{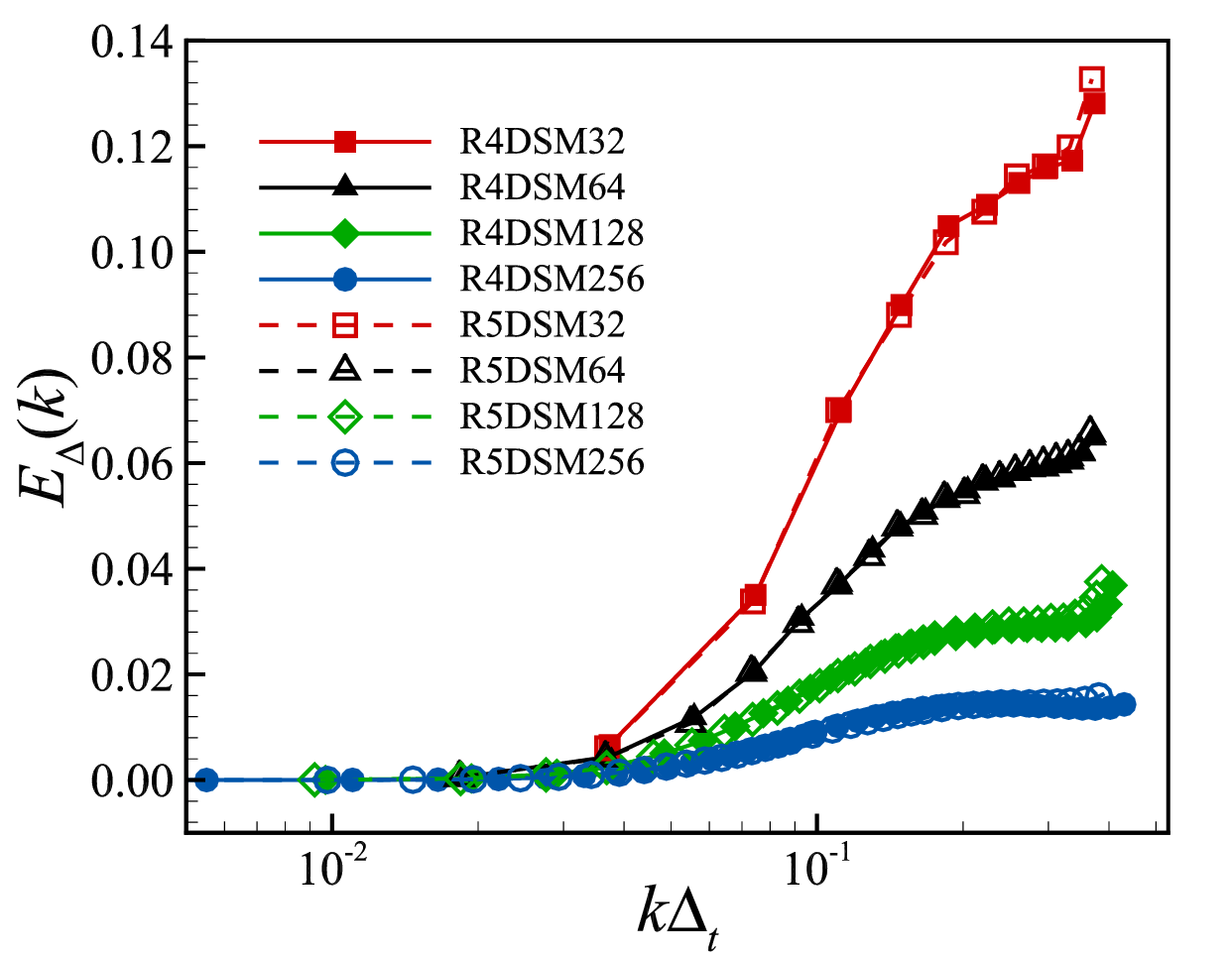}
\ig[width=0.45\lnw]{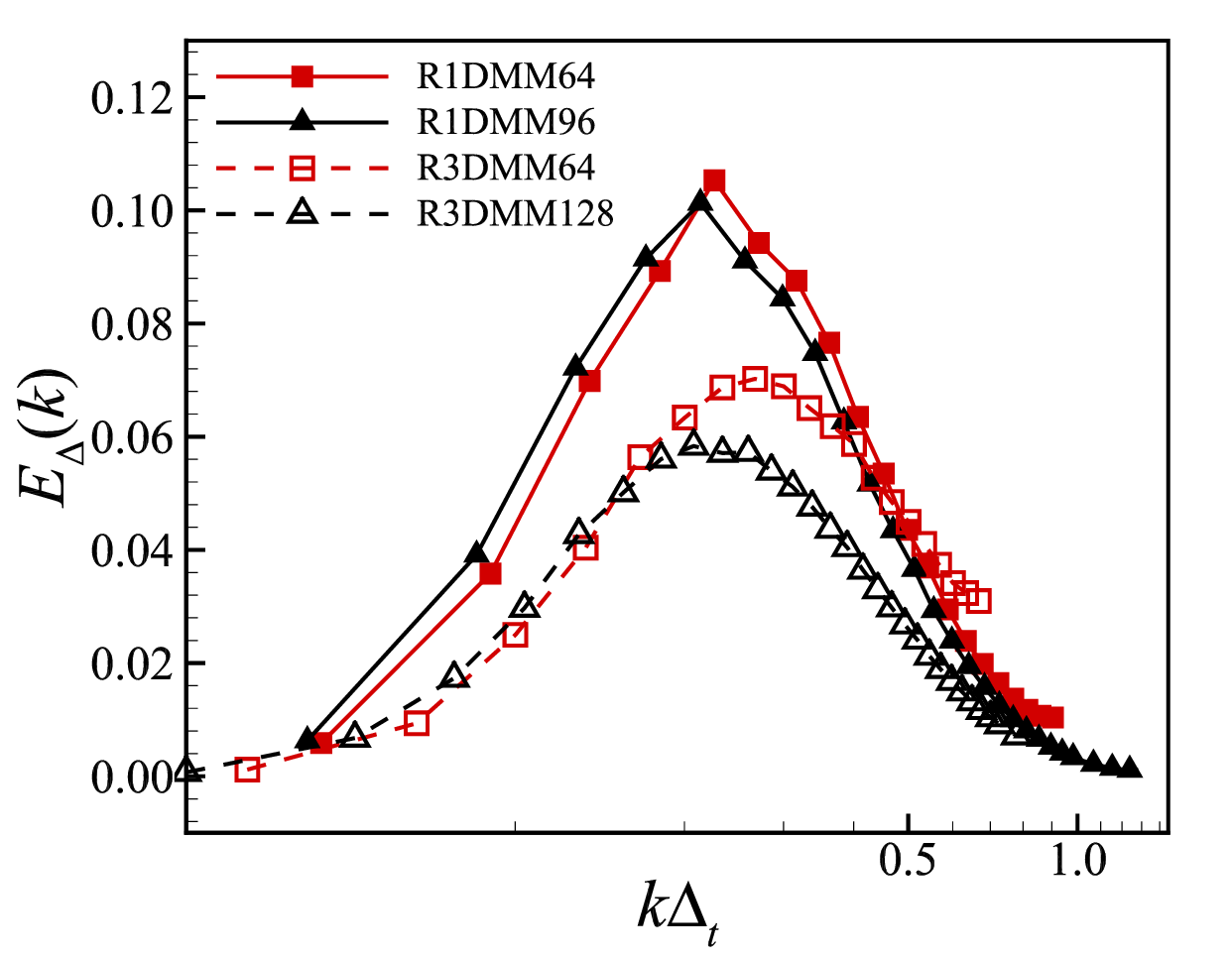}%
\ig[width=0.45\lnw]{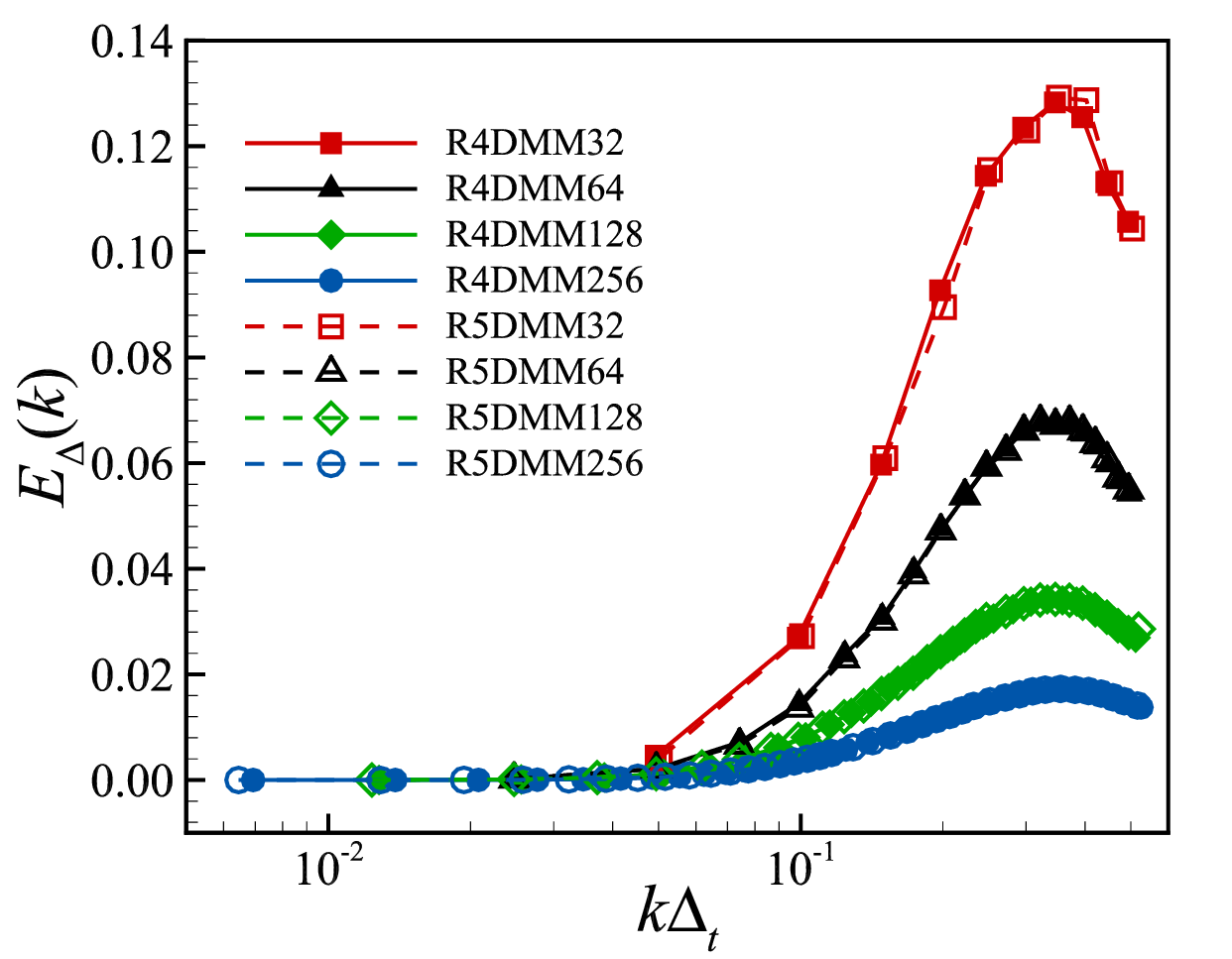}
\caption{\label{fig:llvspec} The energy spectrum $E_\Delta(k)$
for LES with SSM (top), DSM (middle), and DMM (bottom). Results
for lower Reynolds numbers are shown in the left column whereas
results for higher Reynolds numbers are shown in the right
column.}
\efig

The results for $E_\Delta(k)$ calculated from LES data are given
in Fig. \ref{fig:llvspec}. 
Following the normalisation we adopted previously, 
$E_\Delta(k)$ is plotted against $k \Delta_t$, where $\Delta_t$ can be different
for different cases.  
We may comment that, in most cases, 
$E_\Delta(k)$ peaks at an intermediate wavenumber. That is, the perturbations
with energy localised on intermediate wavenumbers are the most
unstable. However, what we are most interested in is the
wavenumber $k_p$ where $E_\Delta(k)$ finds its maximum. 
To compare $k_p$ with $k_c$ quantitatively,
we need to extract the precise values of $k_p$ from 
the spectra. There are
challenges for this since the gap
between two data points can be fairly
large around the peaks. In order to reduce the uncertainty, we use the method in \citet{Lietal24}.
Namely, we fit a smooth curve to the spectrum using
cubic splines. The peak wavenumber of the fitted curve is taken to be $k_p$. 
More details of the method can be found in \citet{Lietal24}.

We consider the results for lower Reynolds numbers first. 
The normalised peak wavenumbers $k_p\Delta_t$ for the LES in subgroups 
R1 to R3
are plotted in the left panel of Fig. \ref{fig:kcetat_LV}
together with the data for $k_c\Delta_t$ in these cases
extracted from Fig. \ref{fig:kcetat}. These data can be
cross-checked with the distributions given in Fig.
\ref{fig:llvspec}.  
There are some discrepancies between the two quantities, 
but the overall
agreement is evident. The discrepancy is somewhat larger for 
DMM at low
$\eta/\Delta$, where $k_p$ is smaller than $k_c$. However, it is
significant to observe that
both $k_p$ and $k_c$ decrease when $\eta/\Delta$ increases,
despite the fact that the two quantities are obtained in 
entirely different ways. 

\bfig
\centering
\ig[width=0.5\lnw]{kcetat_LV_lowRe} 
\ig[width=0.48\lnw]{kcetat_LV_highRe}
\caption{\label{fig:kcetat_LV} Comparison between the
threshold coupling wavenumber $k_c$ and the peak 
wavenumber $k_p$ of the LLV.
Left: data from cases in R1, R2, and R3 (lower Reynolds numbers) 
Right: data from cases in R4 and R5 (high Reynolds numbers), 
and for SSM and DMM only. }
\efig

For the high Reynolds number cases, the physical picture is more
complicated. The peak wavenumber $k_p$ in this case is shown in 
the right panel of Fig. \ref{fig:kcetat_LV}, which corresponds
to  the right column in 
Fig. \ref{fig:llvspec}. As we can see from the latter, for SSM and DMM, 
$E_\Delta(k)$ displays a clear peak at a 
wavenumber that is roughly constant for different Reynolds numbers and 
different $N_\textrm{LES}$. The peak 
wavenumber 
$k_p$ is plotted in the right panel of Fig. \ref{fig:kcetat_LV}. We
can see that the agreement between $k_p$ and $k_c$ is excellent 
for these models over a wide range of values of $\eta/\Delta$. 
However, the results for DSM show unexpected trends. 
On coarse grids, such as in the cases 
with $N_\textrm{LES} = 32$ and $64$, 
$E_\Delta(k)$ for DSM increases with $k \Delta_t$ monotonically. 
Therefore it does not 
peak at an intermediate wavenumber, which means that the 
relationship between $k_p$ and $k_c$ breaks down in these cases. On 
the other hand, it 
is interesting to observe that the middle part of the spectrum appears to be 
amplified relative to the rest of the spectrum when $N_\textrm{LES}$ is increased. This trend can be seen from the diamonds and circles in the middle panel of 
the right column in Fig. \ref{fig:llvspec}. As a matter of fact, 
the spectrum for the case R4DSM256 develops a local maximum 
at $k \Delta_t \approx 0.2$. This value again 
is close to the value for $k_c \Delta_t$ found for the same case. 
This local maximum, however, is not the global maximum. The latter is 
found at the tail of the spectrum. Nevertheless, this observation 
suggests that the same mechanism creating the peak
in $E_\Delta(k)$ for other models is also at play here, 
and a peak may
emerge at higher $N_{\rm LES}$ for the spectrum for DSM too. It would be interesting to
verify this conjecture in the future with larger simulations.

To summarise, we find strong, though not perfect, evidence
to show  
that the relationship between the 
threshold wavenumber for synchronisation and
the peak wavenumber of the LLV also holds in LES. This finding 
adds to 
the evidence already found in DNS of rotating and non-rotating 
turbulence, thus suggesting that the relationship is universal 
among a class of turbulent flows. We further strengthen this claim 
with experiments conducted with the SABRA model below.

\subsection{The synchronisation threshold for the SABRA model}

\bfig
\centering
\ig[width = 0.32\lnw]{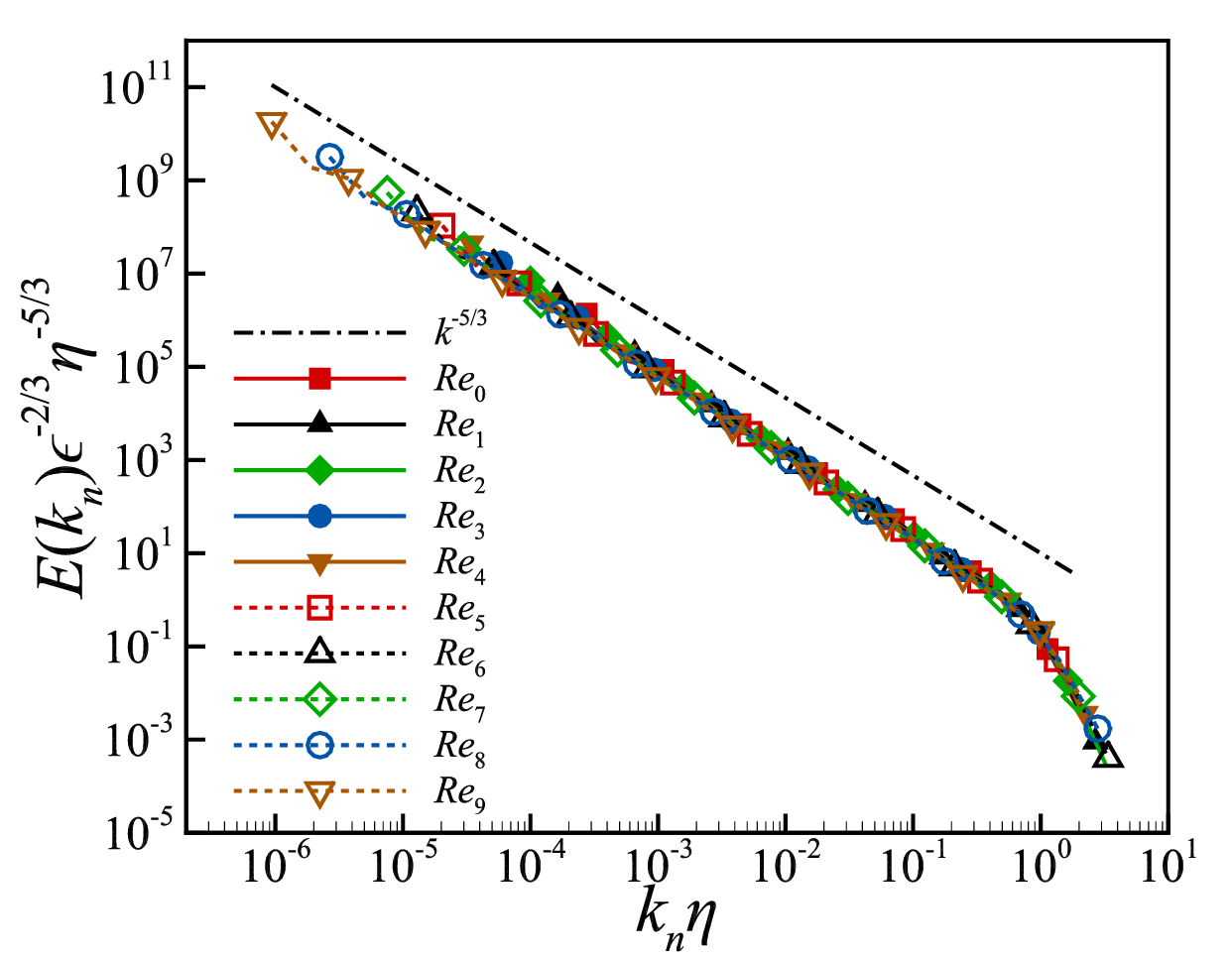} %
\ig[width = 0.32\lnw]{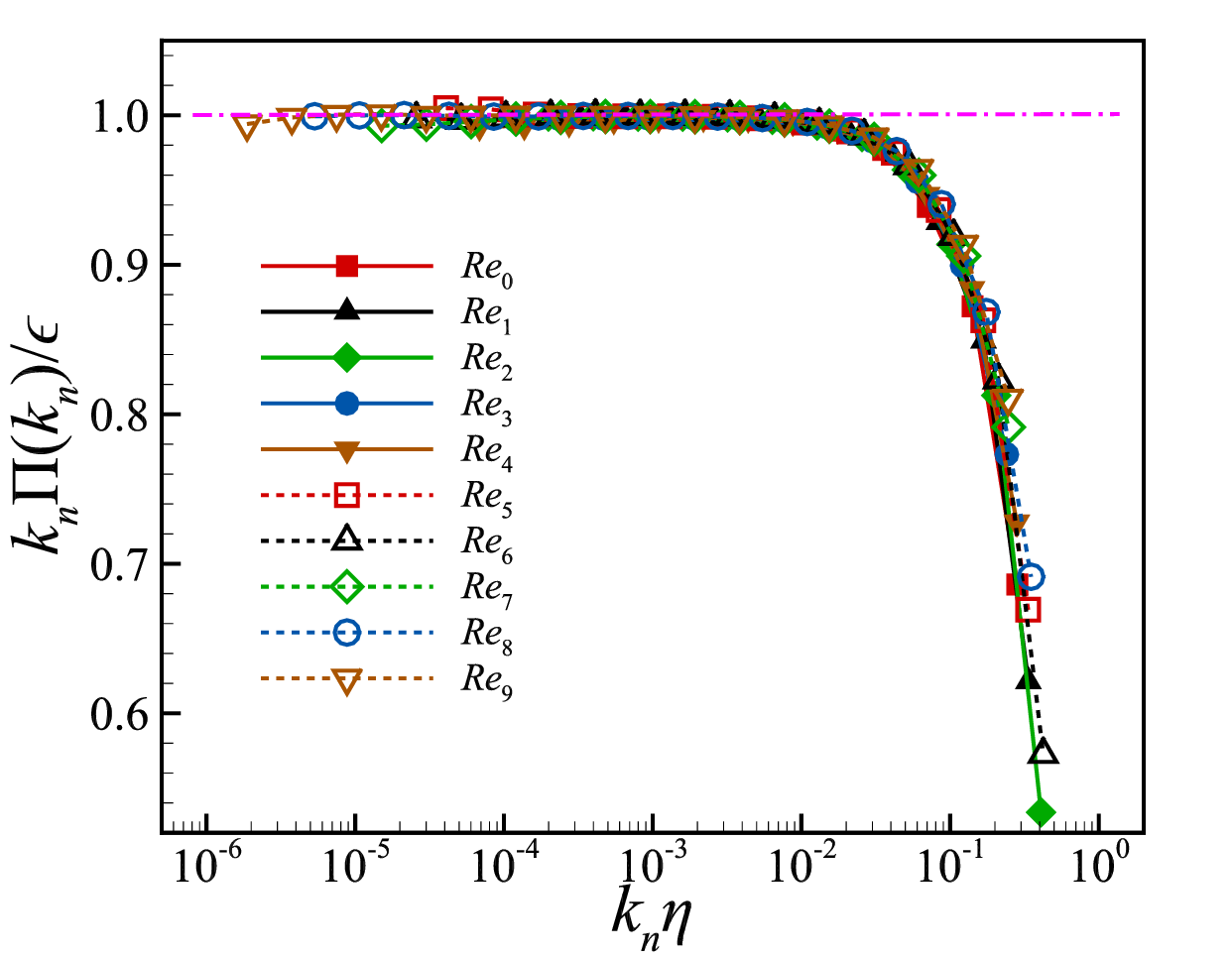} %
\ig[width = 0.32\lnw]{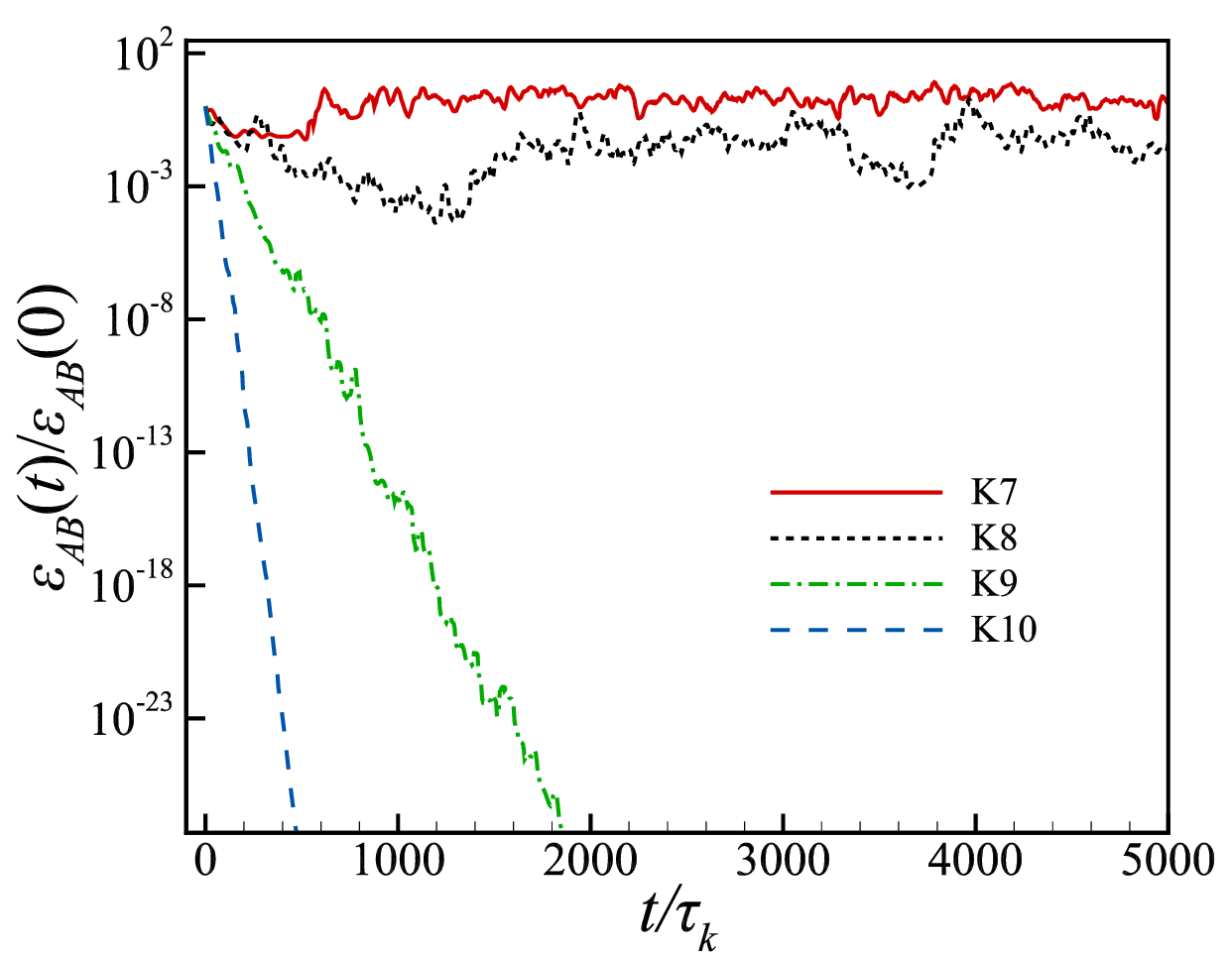} 
\caption{\label{fig:specetc} Left: the (normalised) 
energy spectrum $E(k_n)$. Middle: the (normalised) 
energy flux $\Pi(k_n)$.
Right: the synchronisation error $\EE_{AB }(t)$ for 
case
$Re_0$, with coupling wavenumber 
$k_m = k_7$, $k_8$, $k_9$, and $k_{10}$, respectively.}
\efig

\bfig
\centering
\ig[width = 0.48\lnw]{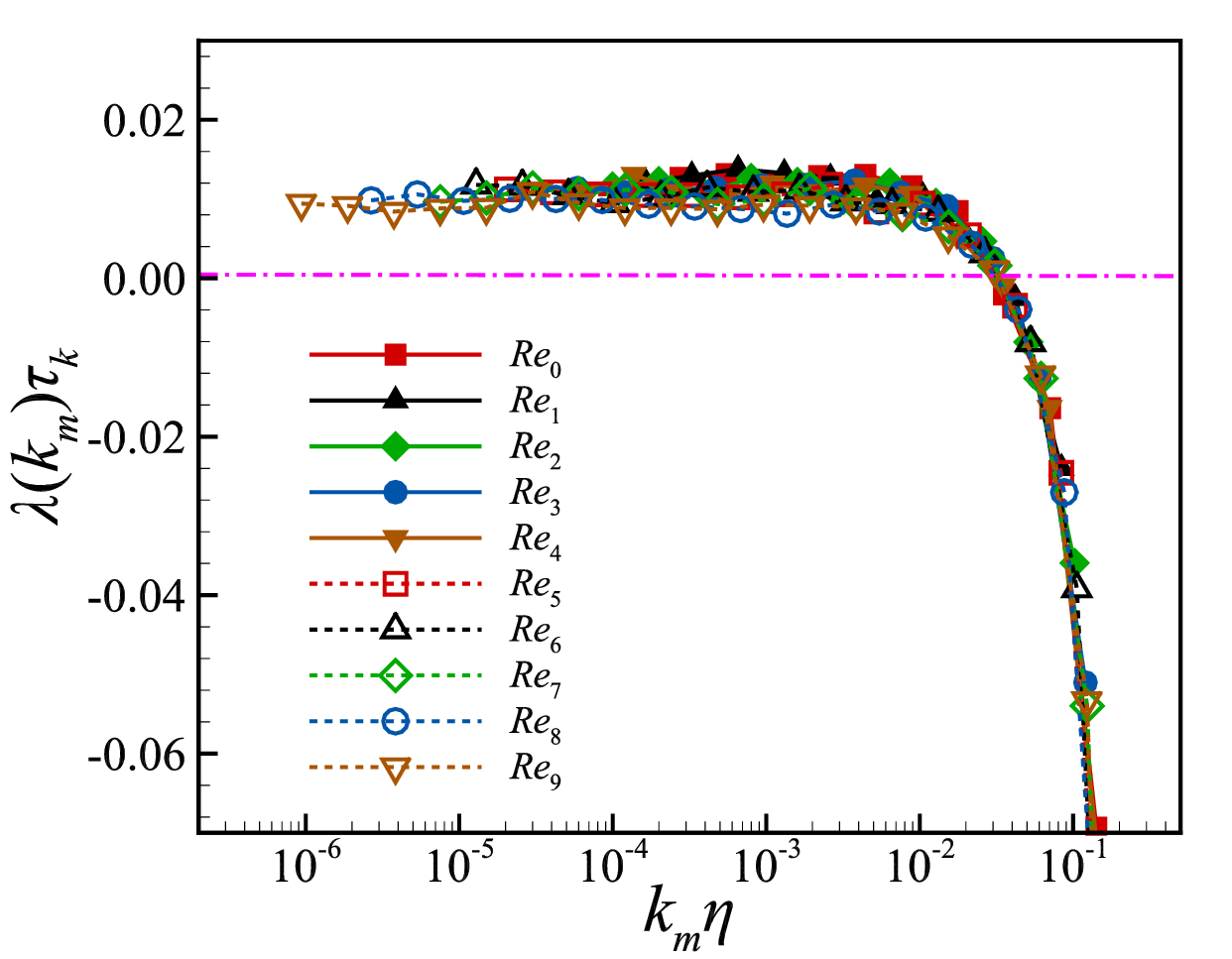}
\ig[width = 0.48\lnw]{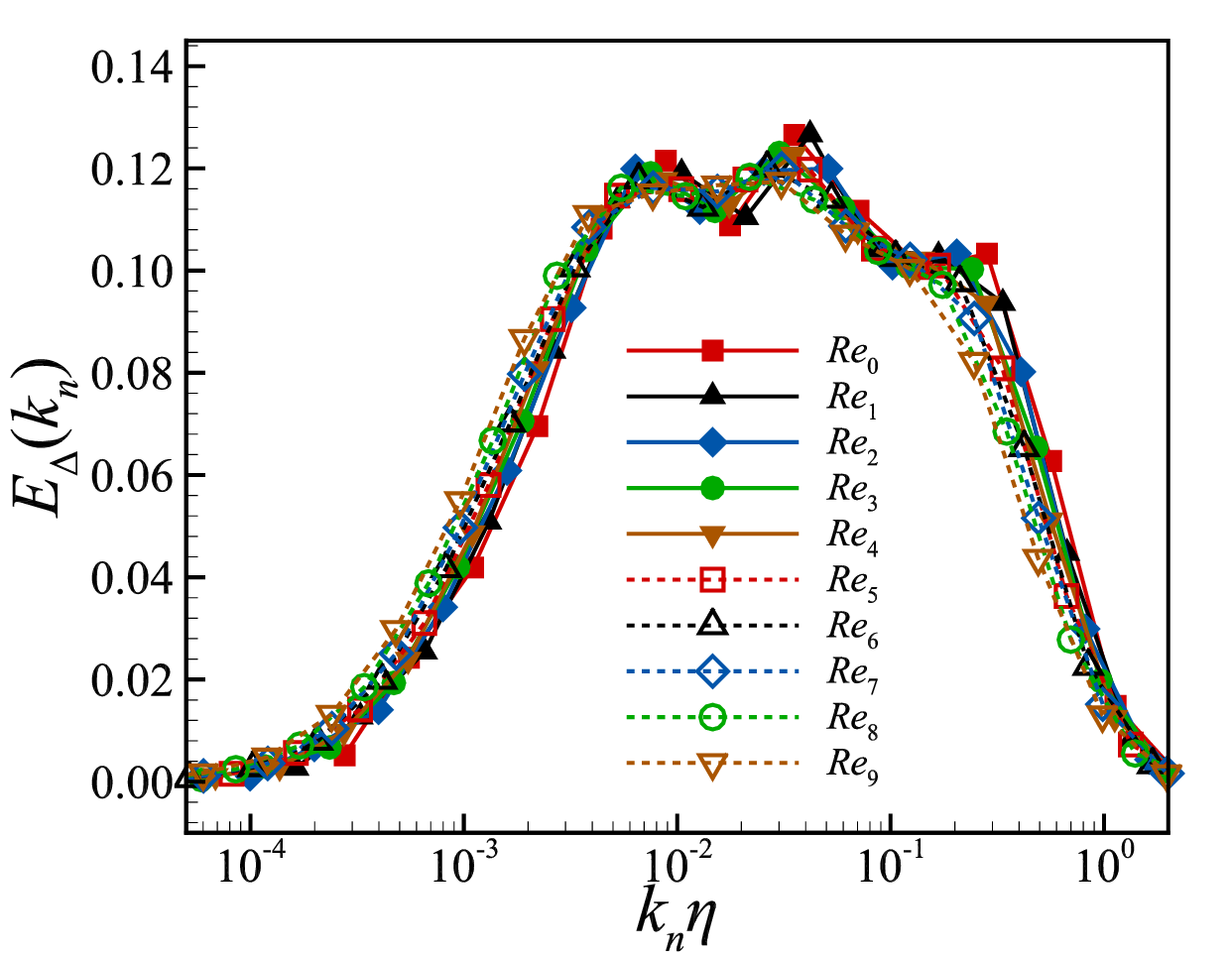}
\caption{\label{fig:sb_cllellv} Left: the normalised conditional 
LLE $\lambda(k_m)$ for the SABRA model. Right: the spectra $E_\Delta(k_n)$
for the LLV.}
\efig

\bfig
\centering
\ig[width = 0.6\lnw]{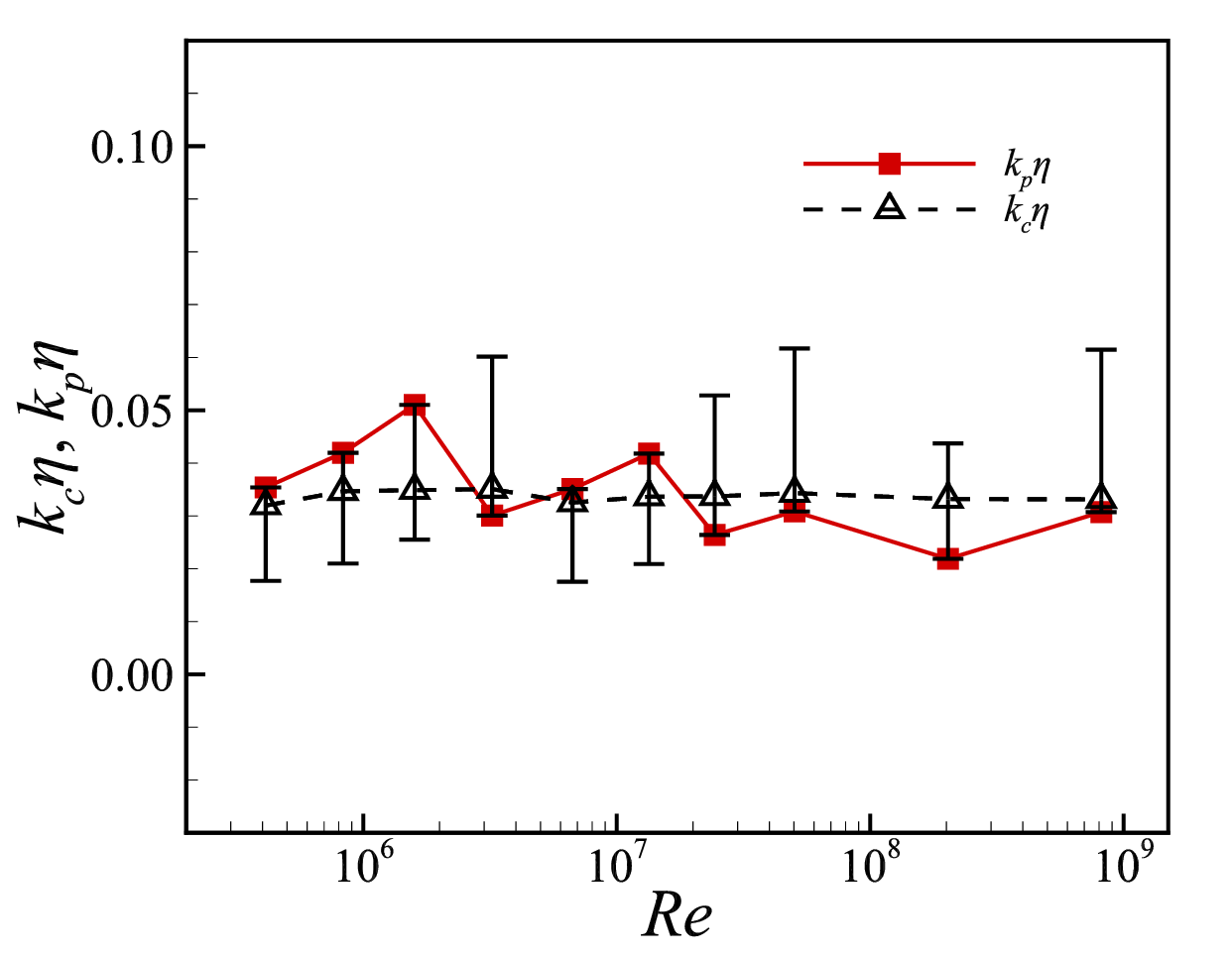}
\caption{\label{fig:sb_kc}The threshold coupling wavenumber $k_c$ obtained from the conditional LLEs $\lambda(k_m )$ versus the peak wavenumber $k_p$ of the energy spectrum of the LLV $E_\Delta(k_n)$.}
\efig

This section focuses on the results of the twin experiments for 
the SABRA model. The computational aspect of the experiments has
been described before. Some additional 
measures that are taken to ensure the accuracy of the results are 
described here. 

When integrating a twin system of the SABRA model, we  
calculate simultaneously the cumulative average of the 
energy spectrum of the LLV $E_\Delta(k_n)$. 
The convergence of $E_\Delta(k_n)$ and 
the conditional LLE $\lambda(k_m)$
is monitored and checked at fixed time intervals. 
Let $t_e$ represent the 
relative change in the core part of $E_\Delta(k_n)$, which, more precisely, is defined as
\begin{equation}
  t_e \equiv \max_n \left\{\left.\frac{\vert \delta E_\Delta(k_n)\vert}{
  E_\Delta (k_n)} ~\right\vert ~E_\Delta(k_n)\ge 10^{-3} \right\},
\end{equation} 
where $\delta E_\Delta(k_n)$ represents the change 
in $E_\Delta(k_n)$ between current and previous checks. Similarly, 
$t_\ell \equiv \vert \delta \lambda(k_m)\vert/\vert \lambda(k_m)\vert$ 
represents the relative 
error in $\lambda(k_m)$ between consecutive checks.
The time integration of the twin systems is conducted 
for at least $9\times 10^7$ time steps, and 
is terminated afterwards if both $t_e$ and $t_\ell$ 
are found to be less than $5 \times 10^{-4}$ for
five consecutive checks. 

The final result for $E_\Delta (k_n)$ is averaged over forty realisations simulated in 
the manner described above. Only one realisation 
is used to calculate $\lambda(k_m)$. As we will see later, 
$E_\Delta(k_n)$ displays 
features unseen in LES (c.f. Fig. \ref{fig:sb_cllellv}). Our additional measures make sure that the 
differences are not numerical artefacts. 
 
We now begin with some basic results about the energy spectrum $E(k_n)$ and the 
energy flux $\Pi(k_n )$. 
$E(k_n)$ is defined by 
\begin{equation}
  E(k_n) = \frac{1}{2} k_n^{-1} 
  \lal \vert u_n\vert^2 \ral.
\end{equation}
For the energy flux, we use a definition suggested by \citet{Biferale03}:
\be 
\Pi(k_n ) = \Im \left\{ \lal u_n^* u_{n+1}^* u_{n+2} \ral
+ \frac{1-b}{q} \lal u_{n-1}^* u_n^* u_{n+1}\ral \right\}
\ee 
where $\Im$ indicates imaginary part, 
$b$ is defined in Eq. (\ref{eq:sb}) and $q$ is the 
ratio between two consecutive wavenumbers. For 
more discussions about these two quantities, including alternative definitions
for $\Pi(k_n)$, see e. g. \citet{Ditlevsen11} and \citet{Biferale03}.

The spectrum, normalised by the parameters given in 
Table \ref{tab:sabra}, is shown in the left panel of 
Fig. \ref{fig:specetc}, for all ten Reynolds numbers. 
The data collapse on one single curve that exhibits a $-5/3$ 
slope over a very wide range of scales. The normalised 
energy flux $\Pi(k_n )$ is shown in the middle panel
of Fig. \ref{fig:specetc}. It plateaus at a value of $1$ for 
$k_n \eta \lesssim 10^{-2}$. This behaviour mimics that of the 
energy flux in energy cascade for 
high Reynolds number turbulence, which is 
expected to be a constant over the scales in the 
inertial ranges. 

The right panel of Fig. \ref{fig:specetc} 
plots the synchronisation error 
$\EE_{AB}(t ) \equiv \Vert {\bm z }_A - \bm z_B\Vert$ for 
the case $Re_0$ for several coupling wavenumber $k_m$, 
which illustrates the general features of the synchronisation process. 
The figure demonstrates that the SABRA model can be synchronised 
in the usual twin experiments. $\EE_{AB}(t)$ 
also decays exponentially for sufficiently large $k_m$, 
but it displays stronger 
fluctuations than what is observed in LES (c.f. Fig. 
\ref{fig:LES_energy_decay_check_R3_64} in Appendix
\ref{sect:e_ab}). Fig. \ref{fig:specetc} shows that synchronisation fails for 
small $k_m$, e.g., for $k_m = k_7$, and  
that the threshold wavenumber $k_c$ is somewhere between $k_8$
and $k_9$ in this case.

Figs. \ref{fig:sb_cllellv} and \ref{fig:sb_kc} are the main
results we intend to obtain from the synchronisation experiments
for the SABRA model. Fig. \ref{fig:sb_cllellv} shows the
conditional LLEs $\lambda(k_m)$ in the left panel, and
$E_\Delta(k_n)$ on the right panel. The threshold wavenumbers
$k_c$ obtained from these two plots are compared in Fig.
\ref{fig:sb_kc}. Before focusing on $k_c$, we comment briefly on
the features of $\lambda(k_m)$ and $E_\Delta(k_n )$. 
The distribution of $\lambda(k_m)$, as can be seen in the left
panel of Fig. \ref{fig:sb_cllellv}, follows the general trend of
the conditional LLEs for LES, with some small differences.
Specifically, $\lambda(k_m)$ is not necessarily a
decreasing function of $k_m$; for small $k_m \eta$, it may
increases with $k_m \eta$. However, $\lambda(k_m)$ eventually becomes negative for
sufficiently large $k_m\eta$, and its distributions seem to
collapse onto one curve as $k_m\eta$ becomes bigger.

The spectrum $E_\Delta(k_n)$, on the other hand, displays more
significant differences. Most importantly, the distributions
might have multiple peaks. For the Reynolds numbers we consider,
there may be one to three peaks. The existence of multiple peaks might be
related to the fact that each shell $u_n$ in the SABRA model is
only coupled to the neighbouring two shells on each side through
the nonlinear term, but we will not explore this further. 

The fact that $E_\Delta(k_n)$ may have multiple peaks indicates
that there might be multiple mechanisms behind the amplification of the 
most unstable perturbation. This observation  
raises the question of whether the maximum of $E_\Delta(k_n)$
plays the same role as the one observed in LES. Nevertheless, we
can still follow our practice with LES results, and identify the
wavenumber $k_p$ where the maximum of $E_\Delta(k_n)$ is located.
Note that $k_p$ has to be one of the discrete wavenumbers $k_n$,
because we do not apply smoothing here when we identify the peak
of $E_\Delta(k_n)$. 
On the other hand, the threshold wavenumber $k_c$ can be
identified from $\lambda(k_m)$, as it is the root of the equation
$\lambda(k_m) = 0$. We find $k_c$ by interpolating between the
two wavenumbers bracketing the root of $\lambda(k_m)$ (c.f., the
left panel of Fig. \ref{fig:sb_cllellv}), so it does not have to
be one of the discrete wavenumber $k_n$. The two nearest
neighbouring wavenumbers are also recorded as indicators for the
uncertainty in determining the threshold wavenumber. 

To test our hypothesis that $k_p$ is the same as the threshold
wavenumber $k_c$, the two wavenumbers are compared in Fig.
\ref{fig:sb_kc}. The two wavenumbers neighbouring $k_c$ are
represented by the error bars in the figure. We observe that $k_c
\eta$ is almost a constant over a wide range of Reynolds numbers,
and its average value is $0.035$. $k_p$ oscillates somewhat
around this value, but it never exceeds the error bars. Actually
it always lands on one of the two error bars. That is, the
difference between $k_p$ and $k_c$ falls within the uncertainty
introduced by the discretisation of the wavenumber $k_n$. This
observation again supports our hypothesis that the peak of the
energy spectrum of the LLV is the same as the synchronisation
threshold wavenumber, which corroborates the observations we made
previously based on LES with SSM and DMM as the SGS stress
models.  

\section{Conclusions \label{sect:con}}

The relationship between the threshold coupling wavenumber 
for two turbulent flows and the leading Lyapunov vector (LLV) for the flows 
is investigated via a series of large eddy simulations (LES)
covering different
subgrid-scale (SGS) stress models, filter scales, 
Reynolds numbers, coupling methods, 
and different
combinations between the master and the slave flows. The large
eddy simulations are complemented by a series of synchronisation 
experiments based on the SABRA model. Our main conclusions can be summarised as
follows.

Firstly, we report a new scaling relation for the leading
Lyapunov exponents (LLE) of LES with canonical SGS stress models.
We introduce the concept of the LLE of filtered DNS, and show
that it displays same scaling behaviour observed in the LLEs of
LES. Phenomenological arguments are presented to give plausible
explanations to  
the empirical data, which thus show that the LLE of LES can be seen as 
approximation to the LLE of the filtered DNS. It appears that
this intuitive interpretation of the LLEs of LES has not been
expounded before. The interpretation strengthens the physical
foundation of the results related to the LLEs of LES reported in
the literature. 

Secondly, we show that synchronisation experiments conducted with
LES and the SABRA model strongly support the hypothesis that, for
a class of chaotic systems, the peak wavenumber of the energy
spectrum of the LLV is the same as the threshold coupling
wavenumber for synchronisation. Evidence shows that DNS of box
turbulence and the SABRA model are among this class. LES for box
turbulence with the standard Smagorinsky model or the dynamic
mixed model also belongs to the class. However, LES with the
dynamic Smagorinsky model generally does not display such a
relationship.

Finally, we show that the conditional LLEs characterise the 
synchronisation process in indirectly coupled systems, as they do
in directly coupled systems. The synchronisation process is also
insensitive to the nature of the master flow. These additional
results help to establish the domain of applicability of our
results. 

The relationship between the threshold coupling wavenumber and the peak
wavenumber of the LLV extends the finding in rotating turbulence reported 
in \citet{Lietal24}, where a qualitative argument 
was also put forward as an explanation for the relationship. The 
argument hypothesizes that the peak of the energy spectrum of the LLV corresponds to the
most unstable Fourier modes in the velocity perturbation, and synchronisation can be
achieved when the growth of these modes is suppressed by coupling. Therefore the
coupling wavenumber should be approximately the same 
as the peak wavenumber of the energy spectrum of the LLV.
We suspect that this argument is similarly applicable here. 
However, an 
analytical theory for the relationship is still 
lacking, and will be the subject of our future research.

%\backsection[Acknowledgments]
%{The authors 
%}

\backsection[Funding]
{Jian Li acknowledges the support of the National Natural Science Foundation of China (No.
12102391).
}

\backsection[Data availability statement]{The data that support the findings of
this study are available from the corresponding author upon reasonable request.}

\backsection[Declaration of Interests]{The authors report no conflict of
interest.}

\appendix

\section{Decay rates of synchronisation error and conditional
Lyapunov exponents \label{sect:e_ab}}

\bfig
\centering
\ig[width=0.5\lnw]{LES_energy_decay_check_R3_64}
\caption{\label{fig:LES_energy_decay_check_R3_64} Comparison between the decay
rates of the synchronisation error $\EE_{AB}(t)$ and the conditional LLEs of
the slave flow.}
\efig

The synchronisation between $\obu_{A}$ and $\obu_{B}$ is
characterised by the synchronisation error 
\be \label{eq:error}
\EE_{AB}(t) = \Vert \obu_{A}-\obu_{B}\Vert.
\ee
Synchronisation happens when $\EE_{AB}(t) \to 0$ as $t\to \infty$. 
Empirically it has been found 
$\EE_{AB}(t) \sim \exp(\alpha t)$, 
i.e., $\EE_{AB}$ decays exponentially when synchronisation happens, 
where $\alpha$ is the decay rate which depends on $k_m$.  
In twin experiments, 
it has been shown \citep{NikolaidisIoannou22, Lietal24} 
that $\alpha$ is equal to the conditional LLE $\lambda(k_m)$. The question 
we address here
is whether they are also the same for three-body experiments. 

Fig. \ref{fig:LES_energy_decay_check_R3_64} shows the results for
$\EE_{AB}(t)$ with symbols for selected three-body experiments. 
It is clear from the figure that 
the decay of $\EE_{AB}(t)$ on average
is exponential for sufficiently large $k_m$. The synchronisation can happen
with very different rates for different models and different $k_m$. 
The lines in the figure represent
exponential decay given by $\exp\left[\lambda(k_m) t\right]$, 
where $\lambda(k_m)$ is the conditional LLE for the slave $\obu_A$. 
The agreement between the symbols and the lines shows 
that the decay rate of the synchronisation error is 
also the same as $\lambda(k_m)$ in these three-body experiments. 
Therefore, the
synchronisation in three-body experiments can also be fully
characterised by the
conditional LLEs.

\section{The impacts of coupling on synchronisation threshold \label{sect:cpl}}

\bfig
\centering
\ig[width=0.48\lnw]{kcetat_different_target_lowRe}
\ig[width=0.48\lnw]{kcetat_different_target_highRe}
\caption{\label{fig:kcetat_diff_target} Threshold wavenumber $k_c \Delta_t$.
Dashed lines with empty symbols: three-body experiments. 
Solid lines with solid symbols:  twin 
experiments (same as those in Fig. \ref{fig:kcetat}). 
 Left: low Reynolds number cases (R1, R2, and R3). 
 Right: high Reynolds number cases (R4 and R5).}
\efig 

We now present data on the threshold wavenumber $k_c \Delta_t$ to show that 
it is insensitive to the coupling mechanism or the master flows. 
Fig. \ref{fig:kcetat_diff_target} compares the thresholds obtained 
from the three-body experiments with those 
obtained from the twin experiments. 
The latter is the same as those shown in Fig. \ref{fig:kcetat}. 
For the data shown in the left
panel with lower Reynolds numbers, the master in 
the three-body experiments is a DNS. 
For the high
Reynolds number data shown in
the right panel, the master is an LES with
a SGS stress model different from the one in the slaves. 
The three cases shown in the figure have a $256^3$ DMM as the master 
and two identical SSM as the slaves, a $256^3$ DMM as the master 
and two identical DSM as the slaves, and a $256^3$ DSM as the master
and two identical DMM as the slaves, respectively. 
For the sake of brevity, we do not conduct 
further tests (e.g., LES with DSM as the master
driving two identical LES with SSM). 

Fig. \ref{fig:kcetat_diff_target} shows that the threshold  
$k_c \Delta_t$ for the twin experiments are very close 
to those obtained in the three-body experiments. This figure is the evidence
for the claim made at the beginning of this subsection. 
Given the close agreement between 
the threshold wavenumbers, it is reasonable to infer that 
the conditional LLEs obtained from the three-body experiments 
would be close to those from the twin experiments too. We omit the 
numerical tests for the latter.

\bibliographystyle{jfm}
\bibliography{turbref}

\begin{thebibliography}{38}
\expandafter\ifx\csname natexlab\endcsname\relax\def\natexlab#1{#1}\fi
\def\au#1{#1} \def\ed#1{#1} \def\yr#1{#1}\def\at#1{#1}\def\jt#1{\textit{#1}}
  \def\bt#1{#1}\def\bvol#1{\textbf{#1}} \def\vol#1{#1} \def\pg#1{#1}
  \def\publ#1{#1}\def\arxiv#1{#1}\def\org#1{#1}\def\st#1{\textit{#1}}

\bibitem[Batchelor(1953)]{Batchelor53}
{\sc \au{Batchelor, G.~K.}} \yr{1953} {\em The theory of homogeneous
  turbulence\/}.  \publ{Cambridge University press, Cambridge}.

\bibitem[Berera \& Ho(2018)]{BereraHo18}
{\sc \au{Berera, A.} \& \au{Ho, R. D. J.~G.}} \yr{2018}  \at{Chaotic properties
  of a turbulent isotropic fluid}.  \jt{Phys. Rev. Lett.}  \bvol{120},
  \pg{024101}.

\bibitem[Biferale(2003)]{Biferale03}
{\sc \au{Biferale, L.}} \yr{2003}  \at{Shell models of energy cascade in
  turbulence}.  \jt{Annu. Rev. Fluid Mech.}  \bvol{35},  \pg{441--468}.

\bibitem[Boccaletti {\em et~al.\/}(2002)Boccaletti, Kurths, Osipov, Valladares
  \& Zhou]{Boccalettietal02}
{\sc \au{Boccaletti, S.}, \au{Kurths, J.}, \au{Osipov, G.}, \au{Valladares,
  D.~L.} \& \au{Zhou, C.~S.}} \yr{2002}  \at{The synchronization of chaotic
  systems}.  \jt{Phys. Rep.}  \bvol{366},  \pg{1--101}.

\bibitem[Boffetta {\em et~al.\/}(2002)Boffetta, Cencini, Falcioni \&
  Vulpiani]{Boffettaetal02a}
{\sc \au{Boffetta, G.}, \au{Cencini, M.}, \au{Falcioni, M.} \& \au{Vulpiani,
  A.}} \yr{2002}  \at{Predictability: a way to characterize complexity}.
  \jt{Phys. Rep.}  \bvol{356},  \pg{367--474}.

\bibitem[Boffetta \& Musacchio(2017)]{BoffettaMusacchio17}
{\sc \au{Boffetta, G.} \& \au{Musacchio, S.}} \yr{2017}  \at{Chaos and
  predictability of homogeneous-isotropic turbulence}.  \jt{Phys. Rev. Lett.}
  \bvol{119},  \pg{054102}.

\bibitem[Bohr {\em et~al.\/}(1998)Bohr, Jensen, Paladin \&
  Vulpiani]{Bohretal98}
{\sc \au{Bohr, T.}, \au{Jensen, M.~H.}, \au{Paladin, G.} \& \au{Vulpiani, A.}}
  \yr{1998} {\em Dynamical Systems Approach to Turbulence\/}.  \publ{Cambridge
  University Press}.

\bibitem[Borue \& Orszag(1996)]{BorueOrszag96}
{\sc \au{Borue, V.} \& \au{Orszag, S.~A.}} \yr{1996}  \at{Numerical study of
  three-dimensional kolmogorov flow at high reynolds numbers}.  \jt{J. Fluid
  Mech.}  \bvol{306},  \pg{293--323}.

\bibitem[Budanur \& Kantz(2022)]{BudanurKantz22}
{\sc \au{Budanur, N.~B.} \& \au{Kantz, H.}} \yr{2022}  \at{Scale-dependent
  error growth in navier-stokes simulations}.  \jt{Phys. Rev. E}  \bvol{106},
  \pg{045102}.

\bibitem[Buzzicotti \& {Di Leoni}(2020)]{BuzzicottiLeoni20}
{\sc \au{Buzzicotti, M.} \& \au{{Di Leoni}, P.~C.}} \yr{2020}
  \at{Synchronizing subgrid scale models of turbulence to data}.  \jt{Phys.
  Fluids}  \bvol{32},  \pg{125116}.

\bibitem[{Di Leoni} {\em et~al.\/}(2018){Di Leoni}, Mazzino \&
  Biferale]{Leonietal18}
{\sc \au{{Di Leoni}, P.~C.}, \au{Mazzino, A.} \& \au{Biferale, L.}} \yr{2018}
  \at{Inferring flow parameters and turbulent configuration with
  physics-informed data assimilation and spectral nudging}.  \jt{Phys. Rev.
  Fluids}  \bvol{3},  \pg{104604}.

\bibitem[{Di Leoni} {\em et~al.\/}(2020){Di Leoni}, Mazzino \&
  Biferale]{Leonietal20}
{\sc \au{{Di Leoni}, P.~C.}, \au{Mazzino, A.} \& \au{Biferale, L.}} \yr{2020}
  \at{Synchronization to big data: Nudging the navier-stokes equations for data
  assimilation of turbulent flows}.  \jt{Phys. Rev. X}  \bvol{10},
  \pg{011023}.

\bibitem[Ditlevsen(2011)]{Ditlevsen11}
{\sc \au{Ditlevsen, P.~D.}} \yr{2011} {\em Turbulence and shell models\/}.
  \publ{Cambridge University press}.

\bibitem[Frisch(1995)]{Frisch95}
{\sc \au{Frisch, U.}} \yr{1995} {\em Turbulence: the legacy of A. N.
  Kolmogorov\/}.  \publ{Cambridge universtiy press, Cambridge}.

\bibitem[Ge {\em et~al.\/}(2023)Ge, Rolland \& Vassilicos]{Geetal23}
{\sc \au{Ge, J.}, \au{Rolland, J.} \& \au{Vassilicos, J.~C.}} \yr{2023}
  \at{The production of uncertainty in three-dimensional navier–stokes
  turbulence}.  \jt{J. Fluid Mech.}  \bvol{977},  \pg{A17}.

\bibitem[Germano(1992)]{Germano92}
{\sc \au{Germano, M.}} \yr{1992}  \at{Turbulence: the filtering approach}.
  \jt{J. Fluid Mech.}  \bvol{238},  \pg{325}.

\bibitem[Goto \& Vassilicos(2009)]{GotoVassilicos09}
{\sc \au{Goto, S.} \& \au{Vassilicos, J.~C.}} \yr{2009}  \at{The dissipation
  rate coefficient of turbulence is not universal and depends on the internal
  stagnation point structure}.  \jt{Phys. Fluids}  \bvol{21},  \pg{035104}.

\bibitem[Henshaw {\em et~al.\/}(2003)Henshaw, Kreiss \&
  Ystr\'{o}m]{Henshawetal03}
{\sc \au{Henshaw, W.D.}, \au{Kreiss, H.-O.} \& \au{Ystr\'{o}m, J.}} \yr{2003}
  \at{Numerical experiments on the interaction between the large- and
  small-scale motions of the navier–stokes equations}.  \jt{Multiscale Model.
  Simul.}  \bvol{1},  \pg{119–149}.

\bibitem[Inubushi {\em et~al.\/}(2023)Inubushi, Saiki, Kobayashi \&
  Goto]{Inubushietal2023}
{\sc \au{Inubushi, M.}, \au{Saiki, Y.}, \au{Kobayashi, M.~U.} \& \au{Goto, S.}}
  \yr{2023}  \at{Characterizing small-scale dynamics of navier-stokes
  turbulence with transverse lyapunov exponents: A data assimilation approach}.
   \jt{Phys. Rev. Lett.}  \bvol{131}~(25),  \pg{254001}.

\bibitem[Kalnay(2003)]{Kalnay03}
{\sc \au{Kalnay, E.}} \yr{2003} {\em Atmospheric modelling, data assimilation
  and predictability\/}.  \publ{Cambridge University press, Cambridge}.

\bibitem[Kang {\em et~al.\/}(2003)Kang, Chester \& Meneveau]{Kangetal03}
{\sc \au{Kang, H.~S.}, \au{Chester, S.} \& \au{Meneveau, C.}} \yr{2003}
  \at{Decaying turbulence in an active-grid-generated flow and comparisons with
  large-eddy simulation}.  \jt{J. Fluid Mech.}  \bvol{480},  \pg{129--160}.

\bibitem[Kuptsov \& Parlitz(2012)]{KuptsovParlitz12}
{\sc \au{Kuptsov, P.~V.} \& \au{Parlitz, U.}} \yr{2012}  \at{Theory and
  computation of covariant lyapunov vectors}.  \jt{J. Nonlinear Sci.}
  \bvol{22},  \pg{727--762}.

\bibitem[Lalescu {\em et~al.\/}(2013)Lalescu, Meneveau \& Eyink]{Lalescuetal13}
{\sc \au{Lalescu, C.~C.}, \au{Meneveau, C.} \& \au{Eyink, G.~L.}} \yr{2013}
  \at{Synchronization of chaos in fully developed turbulence}.  \jt{Phys. Rev.
  Lett.}  \bvol{110},  \pg{084102}.

\bibitem[Li {\em et~al.\/}(2022)Li, Tian \& Li]{Lietal22}
{\sc \au{Li, J.}, \au{Tian, M.} \& \au{Li, Y.}} \yr{2022}  \at{Synchronizing
  large eddy simulations with direct numerical simulations via data
  assimilation}.  \jt{Phys. Fluids}  \bvol{34},  \pg{065108}.

\bibitem[Li {\em et~al.\/}(2024)Li, Tian, Li, Si \& Mohammed]{Lietal24}
{\sc \au{Li, J.}, \au{Tian, M.}, \au{Li, Y.}, \au{Si, W.} \& \au{Mohammed,
  H.~K.}} \yr{2024}  \at{The conditional lyapunov exponents and synchronisation
  of rotating turbulent flows}.  \jt{J. Fluid Mech.}  \bvol{983},  \pg{A1}.

\bibitem[Li {\em et~al.\/}(2020)Li, Zhang, Dong \& Abdullah]{Lietal20}
{\sc \au{Li, Y.}, \au{Zhang, J.}, \au{Dong, G.} \& \au{Abdullah, N.~S.}}
  \yr{2020}  \at{Small-scale reconstruction in three-dimensional kolmogorov
  flows using four-dimensional variational data assimilation}.  \jt{J. Fluid
  Mech.}  \bvol{885},  \pg{A9}.

\bibitem[Meneveau \& Katz(2000)]{MeneveauKatz00}
{\sc \au{Meneveau, C.} \& \au{Katz, J.}} \yr{2000}  \at{Scale-invariance and
  turbulence models for large-eddy simulation}.  \jt{Annu. Rev. Fluid Mech.}
  \bvol{32},  \pg{1--32}.

\bibitem[Nastac {\em et~al.\/}(2017)Nastac, Labahn, Magri \&
  Ihme]{Nastacetal17}
{\sc \au{Nastac, G.}, \au{Labahn, J.~W.}, \au{Magri, L.} \& \au{Ihme, M.}}
  \yr{2017}  \at{Lyapunov exponent as a metric for assessing the dynamic
  content and predictability of large-eddy simulations}.  \jt{Phys. Rev.
  Fluids}  \bvol{2},  \pg{094606}.

\bibitem[Nikitin(2018)]{Nikitin2018}
{\sc \au{Nikitin, N.}} \yr{2018}  \at{Characteristics of the leading lyapunov
  vector in a turbulent channel flow}.  \jt{J. Fluid Mech.}  \bvol{849},
  \pg{942--967}.

\bibitem[Nikolaidis \& Ioannou(2022)]{NikolaidisIoannou22}
{\sc \au{Nikolaidis, M.-A.} \& \au{Ioannou, P.~J.}} \yr{2022}
  \at{Synchronization of low reynolds number plane couette turbulence}.  \jt{J.
  Fluid Mech.}  \bvol{933}.

\bibitem[Pope(2000)]{Pope00}
{\sc \au{Pope, S.~B.}} \yr{2000} {\em Turbulent flows\/}.  \publ{Cambridge
  University Press, Cambridge}.

\bibitem[Sreenivasan \& Antonia(1997)]{SreenivasanAntonia97}
{\sc \au{Sreenivasan, K.~R.} \& \au{Antonia, R.~A.}} \yr{1997}  \at{The
  phenomenology of small-scale turbulence}.  \jt{Annu. Rev. Fluid Mech.}
  \bvol{29},  \pg{435--472}.

\bibitem[Valente \& Vassilicos(2012)]{ValenteVassilicos12}
{\sc \au{Valente, P.} \& \au{Vassilicos, J.~C.}} \yr{2012}  \at{Universal
  dissipation scaling for nonequilibrium turbulence}.  \jt{Phys. Rev. Lett.}
  \bvol{108},  \pg{214503}.

\bibitem[Vassilicos(2015)]{Vassilicos15}
{\sc \au{Vassilicos, J.~Christos}} \yr{2015}  \at{Dissipation in turbulent
  flows}.  \jt{Ann. Rev. Fluid Mech.}  \bvol{47},  \pg{95--114}.

\bibitem[Vela-Martin(2021)]{VelaMartin21}
{\sc \au{Vela-Martin, A.}} \yr{2021}  \at{The synchronisation of intense
  vorticity in isotropic turbulence}.  \jt{J. Fluid Mech.}  \bvol{913},
  \pg{R8}.

\bibitem[Wang \& Zaki(2022)]{WangZaki22}
{\sc \au{Wang, M.} \& \au{Zaki, T.~A.}} \yr{2022}  \at{Synchronization of
  turbulence in channel flow}.  \jt{J. Fluid Mech.}  \bvol{943},  \pg{A4}.

\bibitem[Wang {\em et~al.\/}(2023)Wang, Yuan \& Wang]{Wangetal23}
{\sc \au{Wang, Y.}, \au{Yuan, Z.} \& \au{Wang, J.}} \yr{2023}  \at{{A further
  investigation on the data assimilation-based small-scale reconstruction of
  turbulence}}.  \jt{Phys. Fluids}  \bvol{35}~(1),  \pg{015143}.

\bibitem[Yoshida {\em et~al.\/}(2005)Yoshida, Yamaguchi \&
  Kaneda]{Yoshidaetal05}
{\sc \au{Yoshida, K.}, \au{Yamaguchi, J.} \& \au{Kaneda, Y.}} \yr{2005}
  \at{Regeneration of small eddies by data assimilation in turbulence}.
  \jt{Phys. Rev. Lett.}  \bvol{94},  \pg{014501}.

\end{thebibliography}

\end{document}